\documentclass[a4paper,11pt]{article}
\pdfoutput=1

\usepackage{jheppub-priv}

\hypersetup{
    bookmarksopen,
    pdftitle={Free Energy of a Heavy Quark-Antiquark Pair in a Thermal Medium from AdS/CFT},
    pdfauthor={Carlo Ewerz, Olaf Kaczmarek, Andreas Samberg},
    pdfkeywords={AdS/CFT, holography, temperature, heavy quarks, free energy}
}

\makeatletter
\DeclareRobustCommand*{\bfseries}{%
  \not@math@alphabet\bfseries\mathbf
  \fontseries\bfdefault\selectfont
  \boldmath
}
\makeatother

\input{glyphtounicode}
\newcommand{\MeV}{\,\text{MeV}}

\newcommand{\e}{\text{e}}
\renewcommand{\i}{\text{i}}
\newcommand{\GN}{G_{\text{N}}}
\DeclareMathOperator{\erf}{erf}
\DeclareMathOperator{\erfi}{erfi}

\newcommand{\del}{\partial}
\newcommand{\zt}{z_{\text{t}}}
\newcommand{\zh}{z_{\text{h}}}
\newcommand{\Tc}{T_{\text{c}}}
\newcommand{\AdSfive}{\phantom{}AdS${}_5$}
\newcommand{\LAdS}{L_{\text{AdS}}}
\newcommand{\QbarQ}{Q\bar Q}
\newcommand{\FQQ}{F_{\QbarQ}}
\newcommand{\EQQ}{E_{\QbarQ}}
\newcommand{\SQQ}{S_{\QbarQ}}
\newcommand{\UQQ}{U_{\QbarQ}}
\newcommand{\VQQ}{V_{\QbarQ}}
\newcommand{\fa}{f_{\text{a}}}
\newcommand{\fb}{f_{\text{b}}}
\newcommand{\fc}{f_{\text{c}}}
\newcommand{\fd}{f_{\text{d}}}
\newcommand{\Ls}{L_{\text{s}}}
\newcommand{\Lth}{L_{\text{th}}}
\newcommand{\Lc}{L_{\text{c}}}
\newcommand{\Nc}{N_{\text{c}}}
\newcommand{\SWT}{SW$_T$}
\newcommand{\gYM}{g_{\text{YM}}}
\newcommand{\DeltaSmin}{\Delta S_{\text{min}}}

\renewcommand*\d{\mathop{}\!\mathrm{d}}
\newcommand{\Gammafct}[1]{\operatorname{\Gamma}\left(#1\right)}
\newcommand{\Fhyp}[4]{\operatorname{{}_2F_1}\left(#1,#2;\,#3;\,#4\right)}

\newcommand{\eg}{\textit{e.\,g.}}
\newcommand{\ie}{\textit{i.\,e.}}
\newcommand{\cf}{\textit{cf.}}
\newcommand{\adhoc}{\textit{ad hoc}}

\preprint{BI-TP 2016/05}

\title{Free Energy of a Heavy Quark--Antiquark Pair in a Thermal Medium
  from AdS/CFT}

\author[a,b,c]{Carlo Ewerz,}
\author[d]{Olaf Kaczmarek,}
\author[a,b]{Andreas Samberg}
\affiliation[a]{%
  Institut f\"ur Theoretische Physik,
  Ruprecht-Karls-Universit\"at Heidelberg,\\
  Philosophenweg~16,
  D-69120~Heidelberg, Germany}
\affiliation[b]{%
  ExtreMe Matter Institute EMMI,
  GSI Helmholtzzentrum f\"ur Schwerionenforschung,\\
  Planckstra{\ss}e~1,
  D-64291~Darmstadt, Germany}
\affiliation[c]{Frankfurt Institute for Advanced Studies,\\
Ruth-Moufang-Stra{\ss}e 1, D-60438 Frankfurt, Germany}
\affiliation[d]{%
  Fakult\"at f\"ur Physik,
  Universit\"at Bielefeld,\\
  D-33615 Bielefeld, Germany}

\abstract{%
  We study the free energy of a heavy quark--antiquark pair in a
  thermal medium using the AdS/CFT correspondence.
  We point out that a commonly used prescription for calculating
  this quantity leads to a temperature dependence in conflict
  with general properties of the free energy. The problem originates
  from a particular way of subtracting divergences. We argue that
  the commonly used prescription gives rise to the binding energy
  rather than the free energy.
  We consider a different subtraction procedure 
  and show that the resulting free energy is well-behaved and in qualitative
  agreement with results from lattice QCD. The free energy and the
  binding energy of the quark pair are computed for $\mathcal{N}=4$ 
  supersymmetric Yang--Mills theory and several non-conformal theories. 
  We also calculate the entropy and the internal energy 
  of the pair in these theories. 
  Using the consistent subtraction, we further study the free energy, 
  entropy, and internal energy of a single heavy quark in the thermal medium 
  for various theories. Also here the results are 
  found to be in qualitative agreement with lattice QCD results. 
}

\emailAdd{c.ewerz@thphys.uni-heidelberg.de}
\emailAdd{okacz@physik.uni-bielefeld.de}
\emailAdd{a.samberg@thphys.uni-heidelberg.de}

\begin{document}

\maketitle
\flushbottom

\section{Introduction}
\label{sec:intro}

The AdS/CFT correspondence \cite{Maldacena:1997re,Gubser:1998bc,Witten:1998qj} 
has become a valuable method for studying strongly coupled gauge theories. 
In its original form, the correspondence asserts that string theory of type IIB 
on an anti-de Sitter space, AdS${}_5 \times S_5$, provides an equivalent description of 
a particular conformal field theory, namely 
$\mathcal{N}=4$ supersymmetric $SU(\Nc$) Yang--Mills theory (SYM) 
in four spacetime dimensions. This holographic duality is particularly useful 
in the limit of large $\Nc$ and large 't Hooft coupling $\lambda=\gYM^2\Nc$ in the 
gauge theory, as then the dual description reduces to (super)gravity on AdS${}_5 \times S_5$. 
Remarkably, the weak-coupling limit of the gravity side corresponds to 
the strong-coupling limit of the gauge theory side. A finite temperature $T$ 
of the gauge theory corresponds to a black brane with Hawking temperature $T$ 
in the AdS${}_5$ space. AdS/CFT hence offers an intrinsically nonperturbative 
framework that allows one to study strongly coupled gauge theories both at 
vanishing and at finite temperature. 
The original AdS/CFT correspondence has been extended 
in various ways and by now has found applications to strongly coupled systems 
in many areas of physics. 
One of the most promising applications of the AdS/CFT correspondence 
(or more generally gauge/gravity correspondence) 
concerns the physics of hot, strongly coupled gauge theory plasmas, 
see for example \cite{CasalderreySolana:2011us,DeWolfe:2013cua} for recent reviews. 
The plasmas described in this way are expected to have many properties 
in common with the actual quark--gluon plasma created in heavy-ion collisions, 
and the latter has indeed been found to be strongly coupled at the experimentally 
accessible temperatures somewhat above the critical temperature 
\cite{Arsene:2004fa,Back:2004je,Adams:2005dq,Adcox:2004mh,Aamodt:2010pa,ATLAS:2011ah,Chatrchyan:2012ta}. 

Heavy quarkonia are among the most sensitive probes used in the experimental study 
of the quark--gluon plasma and its properties. Depending on the size of the 
quarkonium state and on the temperature of the plasma, the heavy quark and 
antiquark may be screened from each other, affecting the production rates of 
heavy quarkonia in heavy-ion collisions, see \cite{Matsui:1986dk} for an early, 
influential reference. The static potential of two infinitely heavy quarks and their 
free energy play a central role in the theoretical description of color screening 
and its consequences for heavy quarkonia. 
These two quantities arise naturally in effective field theories which, due to the separation 
of different energy scales, provide a systematic way of dealing with phenomena 
involving heavy quarks both in vacuum and at finite temperature, see 
for instance \cite{Brambilla:2004jw,Brambilla:2008cx}. 
A static pair corresponds to the simplest equilibrium situation one can address 
in this context: a quark and an antiquark, both infinitely heavy, at a given 
distance and being at rest with respect to an infinitely extended plasma. 
Even for this simple situation, the ab initio calculation of the free energy or 
the potential is not a simple task since the presence of a strongly coupled 
plasma requires a nonperturbative framework. Lattice gauge theory has been 
used to calculate the free energy of a static quark pair for QCD 
and quenched versions of it, see for example 
\cite{McLerran:1981pb,Kaczmarek:2002mc,Petreczky:2004pz,Kaczmarek:2004gv,Kaczmarek:2005ui}. 

The AdS/CFT correspondence offers a new and simple method to calculate 
the free energy of a heavy quark--antiquark pair and various related quantities 
in gauge theory plasmas. Besides the simplicity of the calculation, which can 
even be done analytically in some cases, the advantage of the holographic 
description is that it can be applied also to situations which are in general 
prohibitively difficult to attack using lattice gauge theory. Examples 
include dynamical processes with moving quarks, hydrodynamic properties 
of the medium like transport coefficients, or plasmas with a finite 
chemical potential. Although an exact holographic dual of QCD is not known, 
the method promises considerable insight into the dynamics of 
strongly coupled gauge theory plasmas in general, not least because 
many strongly coupled systems appear to share universal features. 
The calculation of the free energy of a static pair in an $\mathcal{N}=4$ SYM plasma 
\cite{Rey:1998bq,Brandhuber:1998bs} was among the first applications 
of the AdS/CFT correspondence. In the meantime, the free energy of a 
heavy quark--antiquark pair has been addressed in the AdS/CFT framework 
for various theories and kinematic situations in different approximations, 
mostly following the ideas and prescriptions of those two early studies. 
The corresponding papers are far too numerous to be cited here comprehensively; 
for a partial list of references see for example \cite{CasalderreySolana:2011us}. 

In the AdS/CFT framework, heavy quarks have a simple description as endpoints 
of macroscopic open strings. These endpoints can move on the 3+1 dimensional 
boundary of the five-dimensional AdS space, which can be identified with 
the physical Minkowski space. 
The string connecting the quark and antiquark hangs down into the bulk, 
that is into the fifth dimension of the AdS space. (The $S_5$ factor in the metric 
does not play an important role for the observables that we consider here.) 
A single heavy quark at rest corresponds to an open string hanging down 
from the boundary into the black hole horizon. 
The dynamics of the quark--antiquark pair or of a single quark is then determined 
by the classical dynamics of the open string in the AdS background, encoded in 
its Nambu--Goto action. 

In the present work, we reconsider the holographic calculation of the free energy 
of the heavy quark--antiquark pair in a thermal medium. 
In particular, we point out an issue with the most commonly used prescriptions 
for the UV renormalization in that calculation. 
Following \cite{Rey:1998bq,Brandhuber:1998bs}, 
the UV divergence in the action of a string connecting a heavy quark pair 
is usually cured by subtracting the action of two strings hanging down into the 
horizon, corresponding to two non-interacting quarks. 
However, this prescription leads to an unphysical temperature dependence of 
the free energy at small distances. The same is true for several other prescriptions 
commonly used in the literature. We will discuss how the temperature 
dependence enters through the corresponding subtraction terms. 
We will then advocate a different, consistent subtraction which avoids this problem. 
Let us note here that essentially equivalent consistent subtractions have been used before
in the finite-temperature context, see\footnote{Although we have thoroughly 
searched the literature that list may be not exhaustive.} 
\cite{BoschiFilho:2006pe,Andreev:2006nw,Noronha:2009ia,Hayata:2012rw,Finazzo:2013aoa,Finazzo:2014zga,Patra:2014qea,Yang:2015aia},  
and the authors of \cite{Bak:2007fk} comment as a side remark that using a
temperature-dependent renormalization is not correct. However, to the
best of our knowledge the full implications of the details of the
renormalization procedure and the distinctions between the quantities
resulting from different procedures have not yet been discussed 
before. Here, we fill this apparent gap in the literature and, in
particular, we study thermodynamic quantities related to the free
energy that crucially require the use of a temperature-independent
renormalization, namely the entropy and the internal energy of the pair. 
We will show that with the consistent UV subtraction the free energy 
as calculated in the AdS/CFT framework is in good qualitative agreement 
with the lattice results. We will also point out that the most commonly used 
subtraction procedure gives rise to the binding energy of the pair rather than 
its free energy, and we will discuss the marked differences between these 
two quantities. As a byproduct of our study we obtain a consistent 
definition of the free energy, the entropy, and the internal energy of 
a single heavy quark in the holographic framework. We will also compare 
the AdS/CFT results for these quantities to lattice results. 

We will consider the observables just described in various theories. 
At finite temperature, despite their different particle content, QCD in the 
deconfined phase and $\mathcal{N}=4$ SYM share some 
essential properties, see for example \cite{CasalderreySolana:2011us}. 
This is in stark contrast to the situation at 
vanishing temperature where these two theories are very  
different. In the following, we will therefore always have the 
quark--gluon plasma phase of QCD in mind when we apply the 
holographic duality. 
But also in the plasma phase, QCD differs from $\mathcal{N}=4$ SYM 
in various respects. The most important difference has its origin in the 
conformal invariance of $\mathcal{N}=4$ SYM. Although the conformality of 
$\mathcal{N}=4$ SYM is broken at finite temperature, the temperature is the only 
dimensionful scale. The behavior of such a plasma at different temperatures 
therefore obeys a scaling of all dimensionful quantities 
with the appropriate powers of temperature.
In QCD, the properties of the plasma depend on temperature in 
a more complicated way due to the presence of other energy scales 
characterizing the dynamics, in particular up to temperatures of 
several times the critical temperature. 
In holography, non-conformal theories closer to QCD can be obtained 
by introducing deformations of pure AdS space. We will in this paper 
only discuss bottom-up approaches of this kind. The simplest way of 
constructing a non-conformal theory is to supplement, in an \adhoc\ way, 
the AdS-black hole metric dual to $\mathcal{N}=4$ SYM with suitable factors, 
often chosen of soft-wall type \cite{Karch:2006pv}, see for example 
\cite{Andreev:2006eh,Kajantie:2006hv}. In general, such a metric does not 
solve the equations of motion of any five-dimensional Einstein--Hilbert action, 
and thus may lead to inconsistencies. A more consistent method is 
to introduce additional scalar fields in the five-dimensional bulk of 
the AdS space. Early examples for this procedure include 
\cite{Csaki:2006ji,Gursoy:2007cb} for vanishing temperature and 
\cite{Gubser:2008ny} for non-vanishing temperature. 
These models can be constructed  
such that the backreaction of the scalar field on the metric changes 
the behavior only in the IR region of the dual field theory, and  
the metric remains asymptotically AdS. One can in fact design 
specific holographic models of strongly coupled QCD along these lines. 
In the present work, instead of considering one specific 
model resembling QCD, we want to take a different approach by studying  
families of consistent non-conformal models. In these models one 
can investigate the general effect of non-conformal deformations and look 
for universal properties common to large classes of non-conformal theories 
at strong coupling. We will indeed find indications for such universal behavior in 
the observables that we study. 

Our paper is organized as follows. 
In sec.~\ref{sec:string-gener-metr} we review the holographic calculation 
of a temporal Wegner--Wilson loop corresponding to a macroscopic string connecting 
a static, heavy quark--antiquark pair in a thermal medium. We use 
a general AdS-type metric that encompasses the holographic dual 
of $\mathcal{N}=4$ SYM and the non-conformal models we want to consider. 
Sec.~\ref{sec:general-discussion} deals with the computation 
of the free energy of the heavy quark--antiquark pair in the 
holographic framework. Here, we discuss in particular the 
temperature dependence of different renormalization prescriptions  
used in this framework and argue that the free energy should be 
independent of temperature at small distances. We then show how 
the free energy and the binding energy are obtained as results of 
different renormalization procedures. 
We illustrate these general considerations for the 
case of $\mathcal{N}=4$ SYM in sec.~\ref{sec:invest-mathc-supersy} 
where we compare the free energy to lattice QCD calculations on the one
hand and to the binding energy on the other hand. 
Next, in sec.~\ref{sec:invest-non-conf}, we investigate how the 
introduction of non-conformality affects the free energy and the binding 
energy of the pair. 
In sec.~\ref{sec:heavy-quark-entropy} we use
the fact that the free energy is a thermodynamic potential and compute
the associated entropy and internal energy for $\mathcal{N}=4$ SYM 
and for the non-conformal models. 
Finally, in sec.~\ref{sec:single-quark-free} we study the free energy and the
associated entropy and internal energy of a single heavy quark in the
strongly coupled plasmas described by our models. In particular, we 
look for universal effects of non-conformal deformations. 
We present our conclusions in sec.~\ref{sec:summary}. Appendix~\ref{sec:deta-comp-qbarq} 
contains some technical details concerning the calculation of the entropy of 
the heavy pair. 

\section{String in general AdS-type metric}
\label{sec:string-gener-metr}

Let us first review the calculation of the string configuration 
holographically corresponding to a heavy quark--antiquark pair in a thermal 
medium. We will do this using a general form of the metric 
of the five-dimensional AdS-type space. The results for particular 
holographic theories, \ie\ $\mathcal{N}=4$ SYM and non-conformal 
models, are later obtained as special cases. 

The most general form of the five-dimensional,
asymptotically AdS metric $g_{MN}$ compatible with translation invariance in the
boundary directions $(t,\vec{x})=(t,x^1,x^2,x^3) = (x^\mu)$ and $SO(3)$
invariance in $\vec{x}$ is given by 
\begin{equation}
\label{eq:metric}
  \d s^2 = \e^{2 A(z)} \left( -h(z)\d t^2 + \d\vec{x}^2 \right) + \frac{\e^{2 B(z)}}{h(z)} \d z^2 \,,
\end{equation}
where $\d s^2 = g_{MN} \d X^M \d X^N$, and $z$ is the fifth-dimensional, holographic coordinate. 

All our model spacetimes are asymptotically AdS. This implies
boundary conditions at $z=0$ for the functions $A$, $B$,
and $h$ in the ansatz \eqref{eq:metric}, namely 
\begin{align}
  A(z) &\sim \log\left(\frac{\LAdS}{z}\right)
         \text{ as }z\rightarrow 0 \,,\label{eq:160}\\
  B(z) &\sim \log\left(\frac{\LAdS}{z}\right)
         \text{ as }z\rightarrow 0 \,,\label{eq:161}\\
  h(z=0) &= 1\,.\label{eq:162}
\end{align}
$\LAdS$ sets the curvature scale of the AdS space. 
A zero in the `blackening' function $h(z)$ signals the presence of a
black hole (more precisely, a black brane extended in the $t$ and $\vec{x}$ 
directions), and we denote its horizon position by $\zh$, that is $h(\zh)=0$. 
For the general metric \eqref{eq:metric} the Hawking temperature 
is \citep{Gubser:2008ny,Gubser:2008yx}
\begin{equation}
  \label{eq:91}
  T = \frac{\e^{A(\zh) - B(\zh)} \lvert h'(\zh) \rvert }{4\pi} \,.
\end{equation}
For simplicity, we will generically refer to asymptotically AdS spaces 
given by metrics with the above properties as `AdS spaces'. 

Already here we point out that the metric \eqref{eq:metric} will 
later be called the Einstein-frame metric. Some of our 
non-conformal models will contain a non-trivial dilaton field 
which affects the coupling of the macroscopic string to the 
background. In those cases the action for the macroscopic string 
needs to be evaluated in the so-called string-frame metric which 
differs from the Einstein-frame metric by a factor involving the 
dilaton, as we will discuss in more detail in 
sec.\ \ref{sec:invest-non-conf}. The calculation of the string 
configuration that we will perform in the rest of the present section 
will assume that the dilaton is absent (or trivial) such that the 
string-frame metric coincides with the Einstein-frame metric. 
It is straightforward to repeat this calculation for a non-trivial 
dilaton as will be needed for some of the non-conformal models 
that we consider in sec.\ \ref{sec:invest-non-conf}. In this context 
we note that the dilaton in our models will always be such that 
also the string-frame metric has asymptotics 
analogous to \eqref{eq:160}--\eqref{eq:162}. More precisely, 
a multiple of the dilaton field will be added to the functions 
$A(z)$ and $B(z)$ in \eqref{eq:metric} without changing their 
leading behavior for $z \to 0$. 

For the quantities that we want to compute, we will need to evaluate the expectation
value of a rectangular Wegner--Wilson loop in the boundary theory.
The holographic calculation of this quantity by means of a dual
macroscopic open string with both endpoints on the boundary
(representing an infinitely heavy quark--antiquark pair) is
well known and can by now be found in textbooks such as
\cite{CasalderreySolana:2011us}. In the following we present 
a brief outline of the calculation, concentrating on the points 
relevant for the discussion in the following sections. 

\begin{figure}[t]
  \centering
  \includegraphics{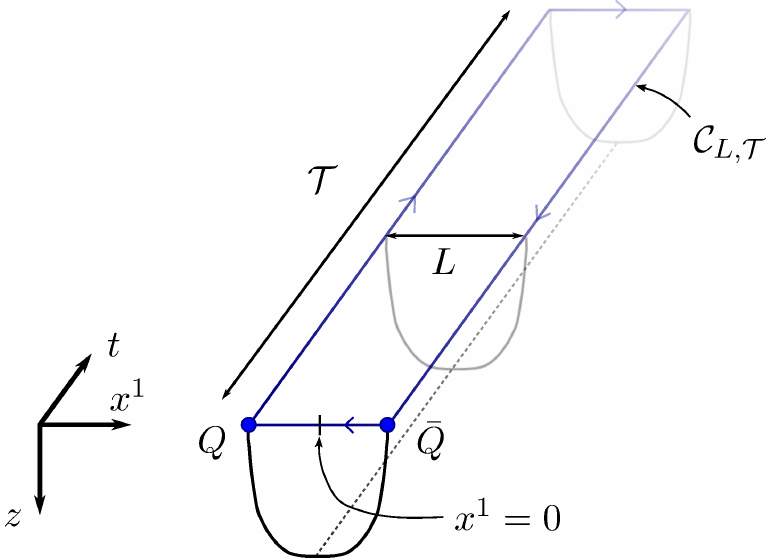}
  \caption{%
    Sketch of a static quark--antiquark pair separated by a distance
    $L$ along the boundary coordinate direction $x^1$. The quarks'
    worldlines are parallel to the time direction $t$, and the
    connecting string extends into the bulk coordinate direction
    $z$. Spacelike slices of the string worldsheet and the integration
    contour $\mathcal{C}_{L,\mathcal{T}}$ used in the integration for
    the Wegner--Wilson loop \eqref{eq:wilsonLoop} are shown. The
    timelike edges of $\mathcal{C}_{L,\mathcal{T}}$ are of length
    $\mathcal{T}$ and coincide with the worldlines of the
    quarks. Eventually, the limit $\mathcal{T}\to\infty$ will be
    taken.%
    \label{fig:sketchWilsonLoopContour}
  }
\end{figure}

The Wegner--Wilson-loop operator in the gauge-theory medium is defined as
\begin{equation}
  \label{eq:wilsonLoop}
    W(\mathcal{C}) = \operatorname{tr} \mathcal{P} \exp\left( \i \oint\limits_{\mathcal{C}} \d{}x^\mu A_\mu(x) \right) \,.
\end{equation}
Here, $\mathcal{C}$ is a closed contour in spacetime,
$A_\mu(x) = A_\mu^a(x) T^a$ is the non-Abelian gauge field where $T^a$ are
the generators in the representation that the trace 
is taken over. $\mathcal{P}$ denotes path ordering.
For our purposes, the integration contour to consider is a rectangular
contour $\mathcal{C}_{L,\mathcal{T}}$, composed of the timelike worldlines of
length $\mathcal{T}$ of the heavy quarks and two small segments along the
spacelike direction in which they are separated by the distance $L$. We choose 
the separation $L$ to be in the $x=x^1$ direction. 
The limit $\mathcal{T} \to \infty$ of infinite temporal extension is required 
for the quantities we want to study. In this limit the contribution from 
the spacelike edges to the integral can be neglected. 
Fig.~\ref{fig:sketchWilsonLoopContour} shows a sketch of this setup.

In the holographic description, the Wegner--Wilson loop is related 
to the Nambu--Goto action of an open string that hangs down into the 
bulk (fifth) dimension of the AdS space and whose endpoints trace out the 
contour of the loop situated at $z=0$. The Nambu--Goto action is 
given by 
\begin{equation}
  \label{eq:42}
  S_{\text{NG}} = - \frac{1}{2\pi\alpha'} \int \d^2\sigma \sqrt{-\det g_{ab}} \,.
\end{equation}
The integral extends over the worldsheet of the string. $1/(2\pi\alpha')$ is the
string tension and $g_{ab}$ the induced metric on the worldsheet,
\begin{equation}
  \label{eq:43}
  g_{ab} = g_{MN} \frac{\partial X^M}{\partial \sigma^a} \frac{\partial X^N}{\partial \sigma^b} \,,\qquad a,b=0,1 \,,
\end{equation}
where $X^M=(t,x,0,0,z(x))$ are the five-dimensional coordinates 
of the string worldsheet in the AdS space. We work in static gauge 
with $\sigma^0=t$ and $\sigma^1=x$. 

In the following we will always assume that the Nambu--Goto action
\eqref{eq:42} captures the relevant dynamics of our problem to a good
approximation.\footnote{Possible corrections to this include thermal 
fluctuations of the string, as addressed for example in 
\cite{Noronha:2009da} or \cite{Grignani:2012iw,Armas:2014nea}.} 
Depending on the model under consideration or on the
approximation to full 10-dimensional string theory there can occur
various corrections, among them couplings of further bulk fields to
the string worldsheet.  Our considerations would then obviously need
to be modified correspondingly. Here, we consider only cases in which
the Nambu--Goto action is a good approximation.

For the general metric \eqref{eq:metric}, we derive the following
explicit form of the Nambu--Goto action \eqref{eq:42} for the string 
worldsheet bounded by $\mathcal{C}_{L,\mathcal{T}}$, 
\begin{equation}
  \label{eq:188}
  S_{\text{NG}} [\mathcal{C}_{L,\mathcal{T}}]
  = -\frac{\mathcal{T}}{2\pi\alpha'}\int_{-L/2}^{L/2}\d{}x\,
  \e^{2A}\sqrt{h\left(1 + \frac{\e^{2B-2A}}{h}\,z'^2\right)} \,,
\end{equation}
where $A$, $B$, and $h$ depend on $z(x)$ and we have performed the
trivial $t$-integration. The function $z(x)$ describes the shape of 
the string hanging into the bulk, and $z'$ denotes the derivative 
with respect to $x$. The endpoints of the string are located at the 
boundary, $z \left(-\frac{L}{2}\right) = z\left(\frac{L}{2}\right) = 0$, and 
the solution is symmetric with respect to $x=0$. 
From the above action one readily derives the equation of motion for the 
embedding $z(x)$, which for $x\ge 0$ gives 
\begin{equation}
  \label{eq:45}
  z'(x) = -\e^{A(z)-B(z)}\sqrt{h(z)\left(\frac{\e^{4A(z)}h(z)}{\e^{4A(\zt)}h(\zt)}-1\right)}\,.
\end{equation}
One finds that the string descends into the bulk, reaching a turning 
point at $z=\zt$ for $x=0$, that is $\zt = z(0)$, 
before symmetrically ascending again towards the boundary. 
This simple string configuration 
is sketched in fig.~\ref{fig:sketchWilsonLoopContour}. 
Using \eqref{eq:45}, the distance $L$ can be expressed as a
function of the turning point $\zt$,
\begin{equation}
  \label{eq:35}
  L(\zt) = 2\int_0^{L/2}\d{}x = 2\int_0^{\zt}\frac{\d{}z}{-z'}
  = 2\int_{0}^{\zt} \d z\, \e^{B-A}
  \left[ h\left(\frac{\e^{4A}h}{\e^{4A_{\text{t}}}h_{\text{t}}} - 1\right)\right]^{-1/2} .
\end{equation}
Here, functions with a subscript `t' are to be evaluated at the
turning point $\zt$.

\begin{figure}[t]
  \centering
  \includegraphics[width=.4\textwidth]{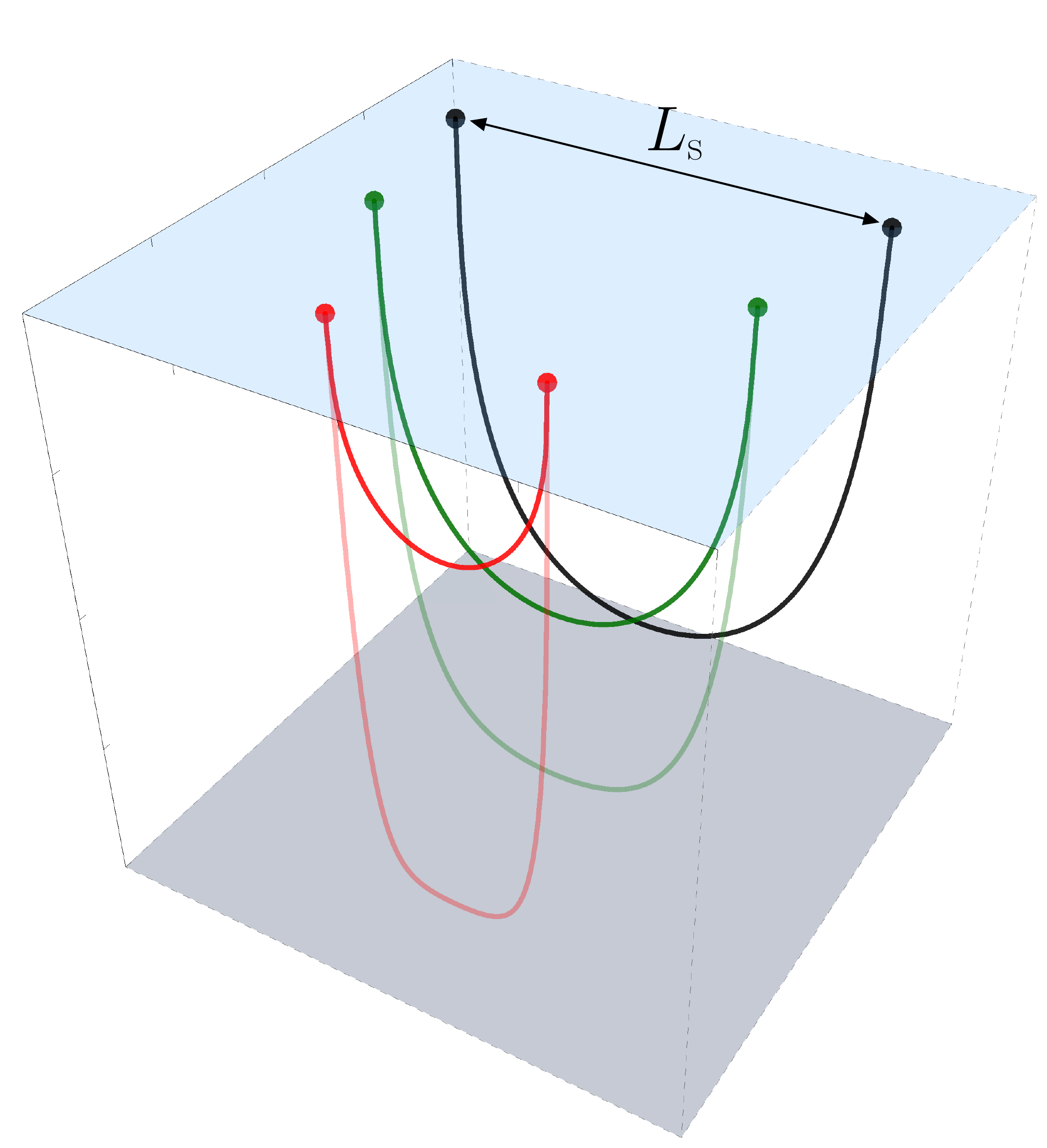}
\hfill 
  \includegraphics[width=.5\textwidth]{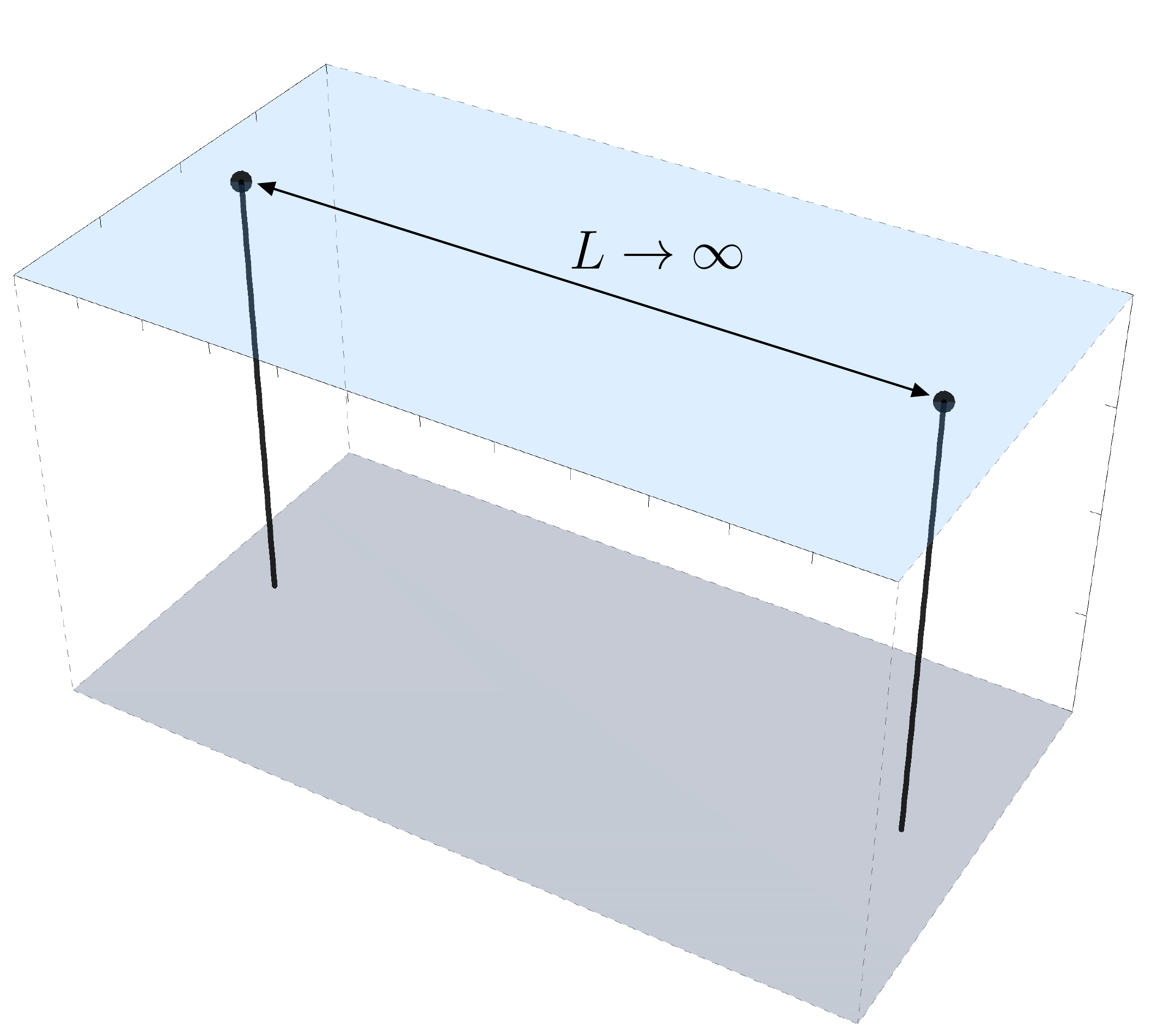}
  \caption{%
   String configurations for the description of a heavy quark--antiquark pair 
   in a thermal medium. The left panel shows string configurations for a 
   bound pair. For each distance $L<\Ls$ there are two string configurations. 
   The one staying further away from the horizon (thick line) is energetically 
   favored over the one protruding further into the bulk (thin line). At the 
   screening distance $\Ls$ the two configurations merge into one. For 
   distances $L>\Ls$ there are no real-valued solutions of this type. The right panel shows 
   the string configuration for a quark and an antiquark at asymptotically large separation, 
   given by two separate straight strings hanging into the black hole. 
   See the text for a discussion of intermediate distances and further string 
   configurations relevant there.%
    \label{fig:stringconfigs3d}
  }
\end{figure}

\begin{figure}[t]
  \centering
  \includegraphics[width=.8\textwidth]{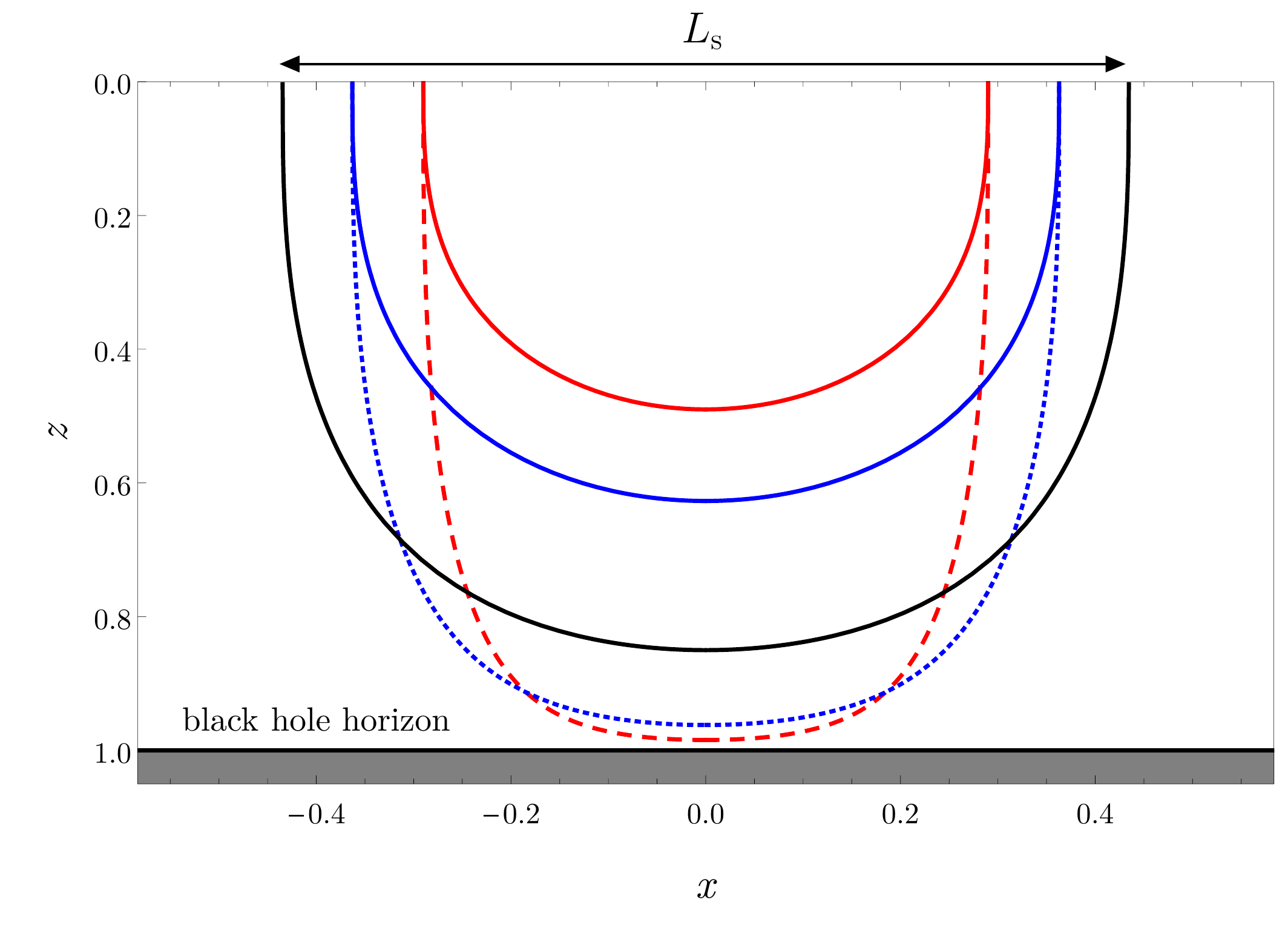}
  \caption{%
   Detailed view of the string configurations for a heavy quark--antiquark 
   pair in a thermal medium of $\mathcal{N}=4$ SYM at distances 
   $L$ below the screening distance $\Ls$. For illustration 
   we have chosen $\zh=1$. The solid lines show the energetically favored 
   configurations, the dashed lines the energetically disfavored configurations.%
    \label{fig:stringconfigs2d}
  }
\end{figure}

As first observed in \cite{Rey:1998bq,Brandhuber:1998bs} for 
the case of $\mathcal{N}=4$ SYM, real-valued solutions of this type 
for the string configuration can only be found for distances $L$ up 
to a certain value $\Ls$ that depends on the temperature. 
The same holds in non-conformal models where the value 
of $\Ls$ then also depends on the chosen metric. The situation 
is qualitatively similar in all models and we want to describe and illustrate it 
now for the $\mathcal{N}=4$ SYM case. The relevant string configurations 
are shown in the left panel of fig.~\ref{fig:stringconfigs3d} and in more 
detail in fig.~\ref{fig:stringconfigs2d}.%
One can uniquely parametrize 
the possible string configurations by their turning point $\zt$.
For each $L$ with $0 \le L < \Ls$ there are two possible string configurations. 
The one staying further away from the black hole horizon turns out 
to have the smaller free energy, while the one coming closer to the 
horizon has a larger free energy and is thus energetically disfavored 
and unstable as we will discuss in more detail in
sec.~\ref{sec:invest-mathc-supersy} below. 
As $L$ is increased the two solutions approach each other and 
at $L=\Ls$ merge into a single solution. None of the solutions 
touches the horizon except for the unstable configuration in 
the limit $L \to 0$ (although the contrary is sometimes claimed 
in the literature). 
In the following, we will concentrate on the energetically favored 
string configurations, but our general considerations will 
apply also to the energetically disfavored configurations. 

$\Ls$ is the maximally possible distance at 
which the quark and antiquark form a bound state connected 
by a string of the type described above. For larger 
distances they become screened by the thermal medium, and we hence call 
$\Ls$ the screening distance. It should not be confused with 
the Debye screening length that characterizes the exponential 
falloff of the interaction between the quark and antiquark at 
still larger distances. The screening distance $\Ls$ can 
in a first approximation be considered as the point where 
the transition from an approximately Coulombic to 
an exponentially damped interaction between 
the quark and antiquark takes place. We will in our calculations 
in the present study only consider inter-quark distances smaller than $\Ls$ 
in the respective models. 

Before we proceed, we want to briefly discuss the behavior 
of the $Q\bar{Q}$ free energy for distances $L > \Ls$. In general, at these distances 
one needs to take into account more complicated string configurations, 
see for example the discussions in \cite{Friess:2006rk,Bak:2007fk}. 
More precisely, the path integral over all string configurations 
connecting the quark and antiquark is dominated by different 
configurations at different inter-quark distances. 
At small distances, as we have discussed before, the dominant 
string configuration is a simple string hanging down into the bulk, 
see the left panel of fig.~\ref{fig:stringconfigs3d}. At asymptotically 
large distances, on the other hand, the dominant string configuration 
is given by two disconnected strings hanging into the black hole, 
see the right panel of fig.~\ref{fig:stringconfigs3d}. Going from 
this configuration towards smaller distances $L$, the two strings 
in the bulk interact via the exchange of certain supergravity modes, 
cf.\ the discussion in \cite{Bak:2007fk}. As elaborated there, 
the mass of the lowest-lying (CT-odd) of these modes is the 
Debye mass $m_{\text{D}}$, which equals the inverse Debye screening length. 
It determines the exponentially attenuated approach of the free energy 
to its asymptotic value at $L \to \infty$. At intermediate but large 
distances, also higher supergravity modes contribute, giving rise 
to a damping of the schematic form $\sum_i \exp(-m_i L)$. 
This behavior has been quantitatively investigated for the case 
of $\mathcal{N}=4$ SYM in \cite{Bak:2007fk}. For non-conformal 
bottom-up models, however, such an analysis appears far more difficult, 
as it would require, among other things, knowledge of the full 
field content of a UV-completed (super)gravity action for the 
respective model. Finally, at distances just above $L_s$, the 
situation appears to be more intricate, as the aforementioned 
configurations below $\Ls$ and the configurations including 
supergravity exchange cannot be matched continuously. 
It appears likely to us that other string configurations 
contribute significantly, see also the related discussion in 
\cite{Friess:2006rk}. Note however, that in \cite{Bak:2007fk} 
it is argued that this transition region has a very small 
extent in $L$ for large 't Hooft coupling $\lambda$. 
In sec.~\ref{sec:invest-mathc-supersy} below we will make 
an observation related to the distinction between the free 
energy and the binding energy which indicates that the 
derivation of that result in \cite{Bak:2007fk} might need to 
be reconsidered. An additional complication related to that 
derivation arises in bottom-up models where the parameter 
$\lambda$ does not necessarily retain its strict interpretation 
as the 't Hooft coupling, as we will see momentarily. 
In view of these difficulties we refrain from making any quantitive 
statements about the region above $\Ls$ in the present study. 
In concluding this discussion for the moment, we should point 
out that also other approaches have been suggested in order 
to treat the region beyond $L_s$, see for example 
\cite{Albacete:2008dz} where the string configuration at $L \le \Ls$ 
is analytically continued to distances above $\Ls$ and becomes 
complex-valued there.

Now we turn back to our main discussion. Below 
we will evaluate the Nambu--Goto action 
\eqref{eq:42} for a macroscopic string propagating in the five-dimensional 
AdS spacetime. It contains the parameter $\alpha'$, and our general metric 
\eqref{eq:metric} always includes a factor $\LAdS^2$ due to
the boundary conditions \eqref{eq:160} and \eqref{eq:161} satisfied by
all our models. We define
\begin{equation}
  \label{eq:190}
  \sqrt\lambda=\frac{\LAdS^2}{\alpha'}
\end{equation}
for the combination of $\LAdS$ and $\alpha'$ that will generically
appear in our observables.
In the holographic dual of $\mathcal{N}=4$ SYM, $\lambda$ coincides 
with the 't Hooft coupling $\lambda=\gYM^2\Nc$. However, when we
consider non-conformal models obtained by non-conformal deformations 
of the bulk theory we have less precise information about the dual 
boundary theory. In particular, we do not know its Lagrangian and 
its (gauge) field content. As a consequence, we cannot be sure of 
the exact meaning of $\lambda$ in the boundary theory. 
In any case, it stands to reason that also in our non-conformal models
$\lambda$ still is a proxy for the coupling strength in the boundary
field theory. 
In phenomenological applications to the quark--gluon plasma one 
would have to dial a particular value for $\lambda$ in a given model 
with a non-conformal deformation. In practice this amounts to treating 
$\lambda$ as an additional free parameter of the model, with the 
caveat that it should be large for the duality to be applicable in the 
approximation that we use. The observables that we consider below 
will always contain a factor $\sqrt\lambda$. 
Since in this work we are mainly interested in their qualitative behavior we will 
divide out that overall factor in plots showing these quantities. 

\section{Free energy versus binding energy of a heavy quark pair}
\label{sec:general-discussion}

We now turn to the heavy quark--antiquark free energy. 
From the field-theory perspective, 
the Wegner--Wilson loop considered in the previous section is a
gauge-invariant object that in particular encodes the free energy of
the $\QbarQ$ pair. To wit, in the limit of infinite temporal extent of
the contour, $\mathcal{T}\to\infty$, we have the relation
\begin{equation}
  \label{eq:33}
  \big\langle W(\mathcal{C}_{L,\mathcal{T}})\big\rangle \sim \exp\left(-\i \FQQ(L) \mathcal{T}\right)
    \,,\qquad \mathcal{T}\rightarrow\infty \,,
\end{equation}
where $\FQQ(L)$ is the $\QbarQ$ free energy
\cite{appelquist1977,Fischler1977,McLerran:1980pk}. 
The expectation value is to be taken for a thermal state of the medium
surrounding the quarks. This introduces the dependence of $\FQQ$ in
\eqref{eq:33} on the temperature $T$ of the medium with which the 
quarks are assumed to be in thermal equilibrium. 
The relation \eqref{eq:33} holds up to an infinite renormalization
constant that we will discuss in the context of the holographic
computation below.

A comment is in order here concerning the distinction between 
the static potential $V_{Q\bar{Q}}$ and the heavy-quark free energy $\FQQ$. 
The former is usually defined via the expectation value of a Wegner--Wilson 
loop as in \eqref{eq:33} above, while the latter is obtained from a corresponding 
expectation value of a Polyakov loop correlator, see for example 
\cite{Burnier:2015tda}.\footnote{More precisely, recent studies starting with
  \cite{Laine:2006ns} have argued that in QCD the real-time
  Wegner--Wilson loop in the limit $\mathcal{T}\to\infty$ gives rise
  to an effective quark potential that is in general complex. However,
  the real part of this potential appears to coincide
  \cite{Beraudo:2007ky} with the $\QbarQ$ (singlet) free energy that
  is defined from a Euclidean-time Wegner--Wilson loop. Indeed, this
  is also observed in lattice QCD calculations which
  reconstruct the real-time potential from the Euclidean-time spectral
  function (see \eg~\cite{Burnier:2014ssa}).}
In our holographic description, we will in the present paper consider the interaction of the 
quarks only for distances $L$ up to the screening distance $\Ls$. 
For this range of distances and in the approximation that we will use, 
the free energy $\FQQ$ of the pair of infinitely heavy 
quarks as computed via \eqref{eq:33} is real-valued and coincides 
with the potential $V_{Q\bar{Q}}$. (This holds up to a temperature-independent 
constant which can be fixed by demanding the zero-temperature limit of 
$\FQQ$ to give the zero-temperature potential $V_{Q\bar{Q}}$, see below.) 
In the present work, we call the quantity extracted from the real-time Wegner--Wilson 
loop via the holographic procedure discussed in the following the $\QbarQ$ 
free energy, and, in accordance with the literature, interpret it as such. 
At distances larger than $L_s$, the free energy $\FQQ$ 
and the potential $V_{Q\bar{Q}}$ will differ, for a recent discussion 
see for example \cite{Burnier:2015nsa}. There, the distinction 
between the two quantities becomes relevant also in the holographic 
description.

Our problem now is to compute the expectation value of the
Wegner--Wilson loop on the gravity side. The basic prescription was
given in \cite{Rey:1998ik,Maldacena:1998im}. From the bulk perspective, the integration contour
$\mathcal{C}$ coincides with the boundary of the worldsheet of the
string dual to the quarks, as discussed in the previous section. 
The expectation value of the Wegner--Wilson loop is then related to
the on-shell string action by
\begin{equation}
  \label{eq:34}
  \big\langle W(\mathcal{C})\big\rangle \sim \exp\left(\i S_{\text{NG}}[\mathcal{C}]\right) \,,
\end{equation}
with $S_{\text{NG}}[\mathcal{C}]$ the extremal Nambu--Goto action of
the string.
This is the saddle-point approximation of the more general statement
where on the right-hand side we would have a path integral over all
string configurations in the bulk with the prescribed boundary
conditions \cite{Maldacena:1998im}.

From equations \eqref{eq:33} and \eqref{eq:34} it follows that the
$\QbarQ$ free energy can be computed holographically from
\begin{equation}
  \label{eq:49}
  \FQQ(L)\sim -\frac{S_{\text{NG}}[\mathcal{C}_{L,\mathcal{T}}]}{\mathcal{T}}\,,
  \qquad \mathcal{T}\to\infty \,.
\end{equation}
This relation still needs to be renormalized. For that we introduce a
regularization on the gravity side as follows. 
For our system, an expression for
$S_{\text{NG}}[\mathcal{C}_{L,\mathcal{T}}]$ can be obtained by
plugging $z'(x)$ from the equation of motion \eqref{eq:45} into the
action functional \eqref{eq:188}. After rewriting the integration over
the coordinate $x$ as an integration over the bulk
coordinate $z$ we obtain 
\begin{equation}
  \label{eq:46}
  S_{\text{NG}}[\mathcal{C}_{L,\mathcal{T}}] = -\frac{\mathcal{T}}{\pi\alpha'}
  \int_{0}^{\zt}\d{}z\,\e^{A+B}
  \sqrt{\frac{\e^{4A}h}{\e^{4A}h-\e^{4A_{\text{t}}}h_{\text{t}}}} \,,
\end{equation}
using again a subscript `t' on functions to indicate their evaluation at the
turning point $\zt$. 
Recall from \eqref{eq:35} that the turning point $\zt$ is directly
related to the $\QbarQ$ distance $L$.
As it stands, the expression \eqref{eq:46} is divergent.
For all models whose metric approaches the AdS metric asymptotically, 
as $z\to 0$, the first factor
$\e^{A+B}$ is asymptotic to $\LAdS^2/z^2$, whereas the square root
approaches unity asymptotically. Thus, we have a divergence from the
lower integral limit, which can be regularized by restricting the
integration to start a small distance $\varepsilon$ away from the
boundary. We thus write for the regularized action 
\begin{equation}
  \label{eq:48}
  S_{\text{NG}}^{\text{(reg)}}[\mathcal{C}_{L,\mathcal{T}}] = -\frac{\mathcal{T}}{\pi\alpha'}
  \int_{\varepsilon}^{\zt}\d{}z\,\e^{A+B}
  \sqrt{\frac{\e^{4A}h}{\e^{4A}h-\e^{4A_{\text{t}}}h_{\text{t}}}}
  \sim -\frac{\mathcal{T}\LAdS^2}{\pi\alpha'}\left(\frac{1}{\varepsilon} + \dots\right)\,.
\end{equation}
The divergence is a pole $\sim 1/\varepsilon$. It
appears because the string endpoints should be situated at the boundary 
$z=0$, which is the holographic realization of the infinite-quark-mass limit
\cite{Maldacena:1998im}.
Subtracting an appropriate (infinite) quantity $\Delta S$ containing 
the $1/\varepsilon$ pole,
we can write \eqref{eq:49} in an operational form for the
computation of the renormalized free energy,
\begin{equation}
  \label{eq:47}
  \FQQ^{\text{(ren)}}(L) = \lim_{\mathcal{T}\to\infty}
  \left(-\frac{S_{\text{NG}}^{\text{(reg)}}[\mathcal{C}_{L,\mathcal{T}}]
      -\Delta S}{\mathcal{T}}\right) \,.
\end{equation}
This expression tacitly includes the limit $\varepsilon\to 0$ that
removes the regulator.
Henceforth, we will drop the specification `ren' and simply write
$\FQQ$ for the renormalized free energy; likewise, we will drop the
superscript `reg'. It remains to specify the subtraction $\Delta S$.

There are two main choices for $\Delta S$ that have been used in the
literature:
\begin{itemize}
\item 
  Expectation values of Wegner--Wilson loops at finite temperature in AdS/CFT 
  were first computed in \cite{Rey:1998bq} and \cite{Brandhuber:1998bs}.  
  There, the subtraction is chosen as twice the action of a straight string
  stretching from the boundary at $z=0$ to the black hole horizon at
  $z=\zh$. This is the commonly used procedure in the literature, it is also
  used in non-conformal theories, see for instance
  \cite{CasalderreySolana:2011us} and references therein.
\item In \cite{Albacete:2008dz} the
  real part of the Nambu--Goto action for infinite $\QbarQ$ distance
  $L$ is subtracted. Given that there are no real solutions to the
  string equation of motion for $L>\Ls$, the authors of
  \cite{Albacete:2008dz} continue the string configuration described above 
  into the complex domain. 
\end{itemize}
These two procedures differ from each other only at non-zero
temperature. For $T\to 0$ both reduce to the procedure used in
the first papers on the computation of the heavy-quark free energy (or
heavy-quark potential) at $T=0$ in AdS/CFT
\cite{Rey:1998ik,Maldacena:1998im}. 

We argue in the following that neither of these procedures is
appropriate for the calculation of the $\QbarQ$ free energy 
at finite temperature. 
Let us first discuss our expectations for this
quantity on the field theory side. For small distances $L$, we expect
the temperature $T$ as well as a possible deformation
scale to have negligible effect on the $\QbarQ$ interaction. 
The physical reason is that the corresponding scales are widely separated: 
the thermal excitations of the medium have typical wavelengths 
of order $1/T$ and hence cannot resolve the interaction of the 
$\QbarQ$ pair at very small distances $L \ll 1/T$. In other words, 
the physics in the UV region of small distances cannot be 
affected by the thermal scale $T$. 
This consideration is supported by data from lattice QCD,
\eg~\cite{Kaczmarek:2004gv,Kaczmarek:2005ui}, where indeed for
$LT\ll 1$ the free energy becomes independent of $T$, see also
\cite{Kaczmarek:2002mc}.
Now consider \eqref{eq:47} for the holographic computation of the
free energy. The first term
$S_{\text{NG}}[\mathcal{C}_{L,\mathcal{T}}]$ becomes independent of
any scale other than $L$ for very small $L$. This is straightforward
to see in the bulk picture for all spacetimes that are asymptotically
AdS, which in particular includes all models that we consider here. 
Note that small $L$ implies a small turning point $\zt$.
(Recall that of the two string configurations
corresponding to a given $L$ we choose the one with the turning
point closer to the boundary, \ie\ the one with smaller $\zt$. We
will explicitly verify below that indeed that string configuration
is energetically preferred over the one with larger $\zt$.)
Thus, a string corresponding to very small $L$ only probes the
part of the spacetime that is essentially fixed by the boundary
conditions and does not depend on the temperature or a possible 
deformation parameter, which manifest themselves
significantly only deeper in the bulk. We will numerically confirm this
bulk argument when discussing the free energy in the following
sections.

Now, if $\FQQ$ should, for small $L$, not depend on $T$ (nor a
potential deformation scale), then also the subtraction $\Delta S$ should not
depend on these scales either. Moreover, $\Delta S$ should not depend
on $L$. 
We therefore advocate a minimal choice $\DeltaSmin$ that just
subtracts the $1/\varepsilon$ pole in the regularized Nambu--Goto
action \eqref{eq:48}. Explicitly, we choose 
\begin{equation}
  \label{eq:50}
  \DeltaSmin \equiv - \frac{\mathcal{T}\LAdS^2}{\pi\alpha'}
             \int_\varepsilon^\infty\frac{\d{}z}{z^2}
  = - \frac{\mathcal{T}\LAdS^2}{\pi\alpha'}\,\frac{1}{\varepsilon} \,.
\end{equation}
As the free energy is defined only up to an overall constant offset, one 
could of course modify $\DeltaSmin$ by an additive constant as long 
as it is $T$-independent, which would correspond to choosing 
a different renormalization scheme. 
Our choice of $\DeltaSmin$ can be used in all models that we will consider here, 
and more generally in any model for which the metric asymptotically 
reduces to AdS.
The choice \eqref{eq:50} ensures that the right-hand side of
\eqref{eq:47} does in fact yield the free energy, and that the
latter does not depend on $T$ (and neither on a possible
deformation scale) for small $\QbarQ$ distances $L$.

Consequently, using formula \eqref{eq:47} with the subtraction
\eqref{eq:50} we find the following expression for the free energy in
terms of a string in our general AdS metric,
\begin{equation}
  \label{eq:51}
  \frac{\pi\FQQ(\zt)}{\sqrt\lambda} = \int_0^{\zt}\d{}z\left[\frac{\e^{A+B}}{\LAdS^2}
    \sqrt{\frac{\e^{4A}h}{\e^{4A}h-\e^{4A_{\text{t}}}h_{\text{t}}}}
    - \frac{1}{z^2}\right] - \frac{1}{\zt} \,.
\end{equation}
Here, we have used the abbreviation $\sqrt\lambda=\LAdS^2/\alpha'$,
see \eqref{eq:190}.
 
Our choice of subtraction \eqref{eq:50} is the holographic 
implementation of what is known to be the correct subtraction 
procedure in field theory. In particular, this subtraction is independent 
of temperature and is determined only by the UV singularity 
(up to an irrelevant $T$-independent constant). 
Let us now consider the problem of choosing the correct subtraction 
from the holographic perspective. Here, a general theory for 
the renormalization of holographic actions has been worked out, 
see \cite{Bianchi:2001kw}. Subtractions are in general encoded 
in covariant local counterterms which can be extracted from the 
near-boundary region. In our case, these counterterms should arise 
from the string action close to the boundary where the background 
metric is arbitrarily close to the pure AdS metric, which in fact 
holds for all finite-$T$ metrics that we consider, see 
\eqref{eq:160}--\eqref{eq:162}. Therefore, the holographic 
counterterms do not depend on the temperature. Neither do they depend 
on the distance $L$ as they are local in the boundary coordinates. 
Both of these conditions are met by our choice $\DeltaSmin$ in \eqref{eq:50}. 
Hence the general holographic counterterm may differ from that choice 
only by a $T$-independent and $L$-independent constant defining 
the renormalization scheme. As we have seen, such 
a constant is not relevant for the free energy. 

In the following we will show that it is indeed possible to express our 
subtraction $\DeltaSmin$ of \eqref{eq:50} in a form exhibiting its 
invariance under boundary transformations induced by bulk diffeomorphisms. 
Similar considerations, although in different contexts, have been 
presented for instance in \cite{Atmaja:2010uu,Taylor:2016aoi}. 
We consider the string configuration illustrated in 
fig.\ \ref{fig:sketchWilsonLoopContour} and treat the descending part 
and the ascending part of the worldsheet separately. Due to the 
symmetry of the string configuration both parts have the same action. 
We can then use a parametrization in terms of $z$ rather than $x$, 
\ie, our worldsheet coordinates are now $z$ and $t$ and we obtain  
$x(z)$ as a function of $z$. 
The Nambu--Goto action for our string worldsheet is 
twice the action of the part descending from $z=0$ to its turning point $\zt$, 
\begin{equation}
\label{eq:add1}
S_{\text{NG}} =  - \frac{1}{2\pi \alpha'} \int \d{}^2\sigma \sqrt{-\det g_{ab}} 
= - \frac{1}{2\pi \alpha'} \, 2 \int_0^{\zt} \d{}z \int_{-\mathcal{T}/2}^{\mathcal{T}/2} 
\d{}t \sqrt{-\det g_{ab}} \,.
\end{equation}
We consider the near-boundary region where we have for the horizon function 
$h(z) \simeq 1$. The bulk metric \eqref{eq:metric} then induces on the 
worldsheet the metric 
\begin{equation}
\label{eq:add2}
\begin{split}
\d s^2_{\text{WS}} &\simeq \frac{\LAdS^2}{z^2} \left( - \d t^2 + (x'^2+1) \d z^2 \right) 
\\
&\simeq \frac{\LAdS^2}{z^2} \left( - \d t^2 + \d z^2 \right) \,. 
\end{split}
\end{equation}
In the second step we have used that close to the boundary $(x'(z))^2\simeq 0$ 
as a consequence of the equation of motion for the string in this parametrization 
due to \eqref{eq:160} and \eqref{eq:161}. (Note that \eqref{eq:add2} also holds 
in the case of a non-trivial dilaton as the corresponding string-frame metric 
also satisfies \eqref{eq:160}--\eqref{eq:162}.) 
We introduce a regularization of the worldsheet by cutting it off at $z=\epsilon$. 
The boundary of the regularized worldsheet is then parametrized by $t$, 
and the induced metric on this worldsheet boundary is 
\begin{equation}
\label{eq:add3}
g_{tt}^{\text{(WS bdry)}}= - \frac{\LAdS^2}{\epsilon^2} \,,
\end{equation}
as we read off from \eqref{eq:add2}. 
We can thus write our subtraction \eqref{eq:50} as 
\begin{equation}
\label{eq:add4}
\begin{split}
\DeltaSmin &= -\frac{\mathcal{T} \LAdS^2}{\pi \alpha'} \int_\epsilon^\infty \frac{\d{}z}{z^2} 
= - \frac{\LAdS^2}{\pi \alpha'} \int_{-\mathcal{T}/2}^{\mathcal{T}/2} \d{}t \int_\epsilon^\infty \frac{\d{}z}{z^2} 
= - \frac{\LAdS}{\pi \alpha'} \int_{-\mathcal{T}/2}^{\mathcal{T}/2} \d{}t \, \frac{\LAdS}{\epsilon}
\\
&= - \frac{\LAdS}{2\pi \alpha'} \, 2 \int_{-\mathcal{T}/2}^{\mathcal{T}/2} \d{}t \, \sqrt{- \det g^{\text{(WS bdry)}}} \,,
\end{split}
\end{equation}
where the determinant in the second line is trivial (\ie\ of a $1 \times 1$ matrix). 
In the limit $\epsilon \to 0$ the worldsheet boundary coincides with 
the contour $\mathcal{C}_{L,\mathcal{T}}$ and we can formally write  
the above expression as 
\begin{equation}
\label{eq:add5}
\DeltaSmin = - \frac{\LAdS}{2\pi \alpha'} \int_{\mathcal{C}_{L,\mathcal{T}}} \d{\xi} \sqrt{- \det g^{\text{(WS bdry)}}} 
\,.
\end{equation}
Here again, for $\mathcal{T} \to \infty$ the spacelike edges at 
timelike infinity do not contribute. 
We have thus found an expression for the subtraction $\DeltaSmin$ which 
in the limit $\mathcal{T} \to \infty$ is manifestly invariant under 
diffeomorphisms $t \to t'(t)$ on the worldsheet boundary. 
This confirms that our choice of subtraction motivated by field theory 
is consistent with the requirements for a counterterm in holographic 
renormalization. 

As we have pointed out in the introductory section, there have 
been studies in the literature using a subtraction procedure 
essentially equivalent to the one that we have described here, see for example 
\cite{BoschiFilho:2006pe,Andreev:2006nw,Noronha:2009ia,Hayata:2012rw,Finazzo:2013aoa,Finazzo:2014zga,Patra:2014qea,Yang:2015aia}. 
However, to the best of our knowledge the consequences of different choices 
of subtraction have not been discussed in depth so far. 
In particular, the different choices for the subtraction give rise to 
different physical quantities as we will describe momentarily. Our aim 
here is to clarify these differences. In the following sections, we will  
then compute further quantities the definition of which crucially 
depends on the correct (temperature-independent) subtraction procedure 
in the computation of the free energy. 

The quantity obtained via the most commonly used subtraction procedure (the
first one in the list above) is the difference of the string action of
the `U'-shaped string connecting the quarks and twice the string
action of a straight string stretching from the boundary to the
horizon,
\begin{equation}
  \label{eq:54}
  \EQQ(L) = \lim_{\mathcal{T}\to\infty}\left({-\frac{S_{\text{NG}}[\mathcal{C}_{L,\mathcal{T}}]
    -2S_{\text{NG}}[\text{straight string}]}{\mathcal{T}}}\right) \,.
\end{equation}
In a general metric of the form \eqref{eq:metric} the Nambu--Goto action for a 
worldsheet corresponding to a static straight string hanging down from the 
boundary to the horizon is given by 
\begin{equation}
  \label{eq:SNGss}
  S_{\text{NG}}[\text{straight string}] = -\frac{\mathcal{T}}{2\pi\alpha'}
  \int_{0}^{\zh}\d{}z\,\e^{A+B} \,.
\end{equation}
It can be regularized in the same way as $ S_{\text{NG}}[\mathcal{C}_{L,\mathcal{T}}]$ 
above by cutting off the integral at a distance $\varepsilon$ away from the boundary. 
Its divergence for $\varepsilon \to 0$ is found in analogy to \eqref{eq:48}, 
\begin{equation}
  \label{regSNGss}
  S_{\text{NG}}^{\text{(reg)}} [\text{straight string}]=
  -\frac{\mathcal{T}}{2\pi\alpha'}
  \int_{\varepsilon}^{\zh}\d{}z\,\e^{A+B} 
  \sim -\frac{\mathcal{T}\LAdS^2}{2\pi\alpha'}\left(\frac{1}{\varepsilon} + \dots\right)\,. 
\end{equation}
Hence, its divergent part equals $\DeltaSmin /2$. We will again drop the 
superscript `reg' in our notation. 

Now the quantity $\EQQ(L)$ in \eqref{eq:54} 
can be understood as a difference of free energies. Namely, by
inserting a zero in the form $-\DeltaSmin+\DeltaSmin$ with the minimal
$\DeltaSmin$ defined in \eqref{eq:50} we can reinterpret this
quantity as the difference of two finite quantities, 
\begin{equation}
  \label{eq:59}
  \begin{split}
    \EQQ(L) &= \lim_{\mathcal{T}\to\infty}\left[{-\frac{\big(S_{\text{NG}}[\mathcal{C}_{L,\mathcal{T}}]
        -\DeltaSmin\big)-\big(2S_{\text{NG}}[\text{straight string}]-\DeltaSmin\big)}{\mathcal{T}}}\right]\\
    &= \FQQ - F_{Q\,;\,\bar Q} \,,
  \end{split}
\end{equation}
where we have used \eqref{eq:47} and have defined 
the free energy of two non-interacting heavy quarks, $F_{Q\,;\,\bar Q}$, as 
\begin{equation}
  \label{FQ-Q}
  F_{Q\,;\,\bar Q} = 
  \lim_{\mathcal{T}\to\infty} \left(-\frac{2S_{\text{NG}}[\text{straight string}]-\DeltaSmin}{\mathcal{T}}\right)\,.
\end{equation}
Due to screening in the hot medium the quark and 
antiquark do not interact at large separation $L \to \infty$. 
Hence one expects $F_{Q\,;\,\bar Q}$ to coincide with the 
large-distance limit of the free energy $F_{Q\bar{Q}}(L)$. 
(Note, however, that the explicit expressions for $F_{Q\bar{Q}}(L)$ given 
above are only valid for $L\le \Ls$ as they are calculated from 
a particular string configuration that exists only in that range.) 
Consequently, we can write $F_{Q\,;\,\bar Q} = 2F_Q$ where we 
may call $F_Q$ the free energy of a single heavy quark, see also \cite{McLerran:1981pb}. 
More explicitly, in our general AdS metric \eqref{eq:metric} we obtain $F_Q$ as 
\begin{equation}
  \label{eq:52}
  \frac{\pi F_Q}{\sqrt\lambda}
  = \frac{1}{2}\left[\int_0^{\zh}\d{}z\left(\frac{\e^{A+B}}{\LAdS^2}-\frac{1}{z^2}\right)
    - \frac{1}{\zh}\right] \,,
\end{equation}
where we have again used the abbreviation
$\sqrt\lambda=\LAdS^2/\alpha'$. We will discuss this single-quark free
energy further in sec.~\ref{sec:single-quark-free}.

Let us turn back to $\EQQ(L)$. We see from \eqref{eq:59} that
$\EQQ(L)$ is an energy difference. It vanishes when the free energy of
the interacting $\QbarQ$ pair equals the free energy of a pair of non-interacting 
heavy quarks. We can thus interpret $\EQQ(L)$ (or more precisely, its negative)
as the binding energy of the $\QbarQ$ pair.
Explicitly, for the binding energy we obtain the relation
\begin{equation}
  \label{eq:39}
  \frac{\pi \EQQ(\zt)}{\sqrt\lambda} =
  \int_{0}^{\zt}\d z\,\frac{\e^{A+B}}{\LAdS^2}
  \left[\sqrt{\frac{\e^{4A}h}{\e^{4A}h-\e^{4A_{\text{t}}}h_{\text{t}}}} - 1\right]
  - \int_{\zt}^{\zh}\d z\,\frac{\e^{A+B}}{\LAdS^2} \,.
\end{equation}

The binding energy has been extensively studied (often as a `finite-temperature
quark--antiquark potential') by means of the gauge/gravity duality, see
for instance
\cite{Rey:1998bq,Brandhuber:1998bs,Kinar:1998vq,BoschiFilho:2006pe,
  Liu:2006he,Liu:2008tz,He:2010ye,Fadafan:2011gm,Ewerz:2012fca}. These
references include investigations in $\mathcal{N}=4$ SYM (in the
strict limit of infinite 't Hooft coupling $\lambda$ as well as
including first-order corrections in an expansion in $1/\lambda$
\cite{Fadafan:2011gm}) and in models with non-conformal deformation, 
at vanishing and non-zero temperature, and with the $\QbarQ$ pair stationary 
or moving with respect to the rest frame of the background medium (including
analyses of the dependence on the angle of the
$\QbarQ$ dipole to its velocity \cite{Liu:2006he}).
Furthermore, $\EQQ$ has been studied in holographic models of
anisotropic strongly coupled plasma
\cite{Giataganas:2012zy,Rebhan:2012bw}, as well as at
non-zero chemical potential in $\mathcal{N}=4$ SYM
\cite{Caceres:2006ta} and non-conformal models \cite{Ewerz:2013wva}.

We will see in the following sections that the behavior of the binding energy
is fundamentally different from that of the free energy. Moreover, we
will find that the free energy in $\mathcal{N}=4$ SYM, as in the
non-conformal models, behaves qualitatively like the $\QbarQ$ free
energy computed in lattice QCD, whereas the binding energy does not.
This corroborates our general arguments for the use of the subtraction
\eqref{eq:50} for the computation of the free energy.

\section{Free energy and binding energy in
  \texorpdfstring{$\mathcal{N}=4$ supersymmetric Yang--Mills}{N = 4 supersymmetric Yang-Mills}
  theory}
\label{sec:invest-mathc-supersy}

In the previous section we have obtained expressions for the $\QbarQ$ free energy
$\FQQ$ and binding energy $\EQQ$ in general holographic models, see
\eqref{eq:51} and \eqref{eq:39}, respectively. 
Now we want to study the properties of these two quantities for the simplest case, 
namely $\mathcal{N}=4$ SYM whose gravity dual is given by pure AdS-black hole space. 
We will find the general behavior of the free energy to be in qualitative agreement 
with the behavior found in lattice QCD. 

As is well known, $\mathcal{N}=4$ supersymmetric Yang--Mills 
theory at non-zero temperature in the limits of large number of colors
and large 't Hooft coupling can be described by a
classical gravity theory with the \AdSfive{}-Schwarzschild metric
\begin{equation}
  \label{eq:78}
  \begin{aligned}
    \d{}s^2 &= \frac{\LAdS^2}{z^2} \left( -h(z)\d{}t^2 + \d{}\vec{x}^2
      + \frac{1}{h(z)}\d{}z^2 \right) \,,\\
    h(z) &= 1 - \frac{z^4}{\zh^4} \,.
  \end{aligned}
\end{equation}
In other words, for $\mathcal{N}=4$ SYM the functions $A$ and $B$ in the 
general metric \eqref{eq:metric} are given by $A(z)=B(z)= \log (\LAdS/z)$. 
At $z=\zh$ there is a planar black-hole (black-brane) horizon whose
associated Hawking temperature, 
\begin{equation}
  \label{eq:82}
  T = \frac{1}{\pi\zh} \,,
\end{equation}
is identified with the temperature of the boundary field theory.

In $\mathcal{N}=4$ SYM, we can evaluate the integrals in the
formulae for $\FQQ$, $\EQQ$, and $L$ explicitly.
Using \eqref{eq:39} for the binding energy $\EQQ(L)$ with the
AdS-Schwarz\-schild metric \eqref{eq:78}, we obtain
\begin{equation}
  \label{eq:60}
  \frac{\pi\EQQ(\zt)}{\sqrt\lambda}
  = -\frac{\sqrt\pi\Gammafct{\frac{3}{4}}}{\Gammafct{\frac{1}{4}}}
  \Fhyp{-\frac{1}{2}}{-\frac{1}{4}}{\frac{1}{4}}{\frac{\zt^4}{\zh^4}}
  \frac{1}{\zt} + \frac{1}{\zh} \,,
\end{equation}
where $\operatorname{{}_2F_1}$ is the (Gau\ss{}ian) hypergeometric
function. An equivalent formula has been obtained in
\cite{Avramis:2006em}. Note that, working in $\mathcal{N}=4$ SYM,
$\sqrt\lambda$ which we defined as a shorthand for the ratio of bulk
quantities $\LAdS^2/\alpha'$ is in fact the 't Hooft coupling
$\sqrt\lambda=\gYM^2\Nc$. 

The temperature-dependent term $1/\zh$ in \eqref{eq:60} is
entirely due to the contribution
$2S_{\text{NG}}[\text{straight string}]$ of the straight strings
stretching from the boundary to the horizon, see \eqref{eq:54}.
It is due to this term that $\EQQ$ depends on $T$ for small inter-quark
distances. As discussed above, this should not be the case
for the free energy $\FQQ$, and $\FQQ$ indeed lacks that term.
Explicitly, we find from \eqref{eq:51} with the metric \eqref{eq:78} 
\begin{equation}
  \label{eq:61}
  \frac{\pi\FQQ(\zt)}{\sqrt\lambda}
  = -\frac{\sqrt\pi\Gammafct{\frac{3}{4}}}{\Gammafct{\frac{1}{4}}}
  \Fhyp{-\frac{1}{2}}{-\frac{1}{4}}{\frac{1}{4}}{\frac{\zt^4}{\zh^4}}
  \frac{1}{\zt} \,.
\end{equation}
Comparing this to the binding energy $\EQQ$ 
we find that $\FQQ(L) < \EQQ(L)$ for any $T>0$, as $\EQQ$ in \eqref{eq:60} 
has an additional positive contribution $1/\zh$ (and $L$ is in one-to-one 
correspondence to $\zt$). 
In the limit $T\to 0$, which is $\zh\to\infty$
on the gravity side, $\FQQ$ and $\EQQ$ coincide.

We have expressed both the binding energy and the free energy in terms of
the turning point $\zt$. The latter is related to the inter-quark
distance $L$ via the explicit relation
\begin{equation}
  \label{eq:62}
  L(\zt) = \frac{2\sqrt\pi\Gammafct{\frac{7}{4}}}{3\Gammafct{\frac{5}{4}}}
  \sqrt{1-\frac{\zt^4}{\zh^4}}
  \Fhyp{\frac{1}{2}}{\frac{3}{4}}{\frac{5}{4}}{\frac{\zt^4}{\zh^4}}\zt \,,
\end{equation}
derived from the general expression \eqref{eq:35}. This explicit form
has also been obtained in \cite{Avramis:2006em}.

For $T=0$ it is possible
to explicitly solve \eqref{eq:62} for $\zt$ and to compute
$\VQQ(L)\equiv\FQQ(L)=\EQQ(L)$, the heavy quark--antiquark potential,
analytically as a function of $L$,
\begin{equation}
  \label{eq:77}
  \VQQ(L) = -\frac{4\pi^2\sqrt\lambda}{{\Gamma}^4\!\left(\frac{1}{4}\right) L} \,,
\end{equation}
which has been first obtained in \cite{Maldacena:1998im}. 
The strict proportionality $\VQQ\propto 1/L$
reflects the absence of any other dimensionful scale at $T=0$.

There is another interesting fact that we would like to point out here. 
In $\mathcal{N}=4$ SYM, the zero-temperature limit 
of the action \eqref{regSNGss} for a straight string stretching from
the boundary to the horizon is, as a consequence of \eqref{eq:78} and
\eqref{eq:82},
\begin{equation}
\label{N4ssT0}
S_{\text{NG}}[\text{straight string}]\big|_{\mathcal{N}=4,T= 0}
= - \frac{\mathcal{T}\LAdS^2}{2\pi\alpha'}
             \int_\varepsilon^\infty\frac{\d{}z}{z^2} \,.
\end{equation}
This quantity receives a contribution only from the lower limit of 
the integral and is thus fully given by the UV divergence of the straight string. 
It exactly coincides with one half of the minimal subtraction $\DeltaSmin$ 
that we advocate for the definition of $\FQQ$ in all models and 
for all temperatures, see \eqref{eq:50}. One could therefore think 
of the minimal subtraction for any asymptotically AdS metric 
as subtracting the action of two straight strings corresponding to 
pure $\mathcal{N}=4$ SYM at $T=0$.\footnote{In this context, we 
should mention that another possible subtraction for the case of 
general asymptotically AdS spaces would be the action of two 
straight strings in the $T \to 0$ limit of the corresponding theory. 
Such a subtraction would be given by \eqref{regSNGss} with 
$\zh \to \infty$ and the functions $A$ and $B$ replaced by their 
$T=0$ forms. This prescription would differ from our minimal 
subtraction procedure only by a $T$- and $L$-independent 
constant, and thus correspond to a different renormalization 
scheme. However, the limit of vanishing temperature might be 
difficult to take for some of the bottom-up models proposed in 
the literature, for example for models that are constructed 
aiming mainly at a description of the high-temperature phase.  
Some of these models have an accessible range in temperature $T$ 
which is limited from below. The minimal subtraction that we 
advocate here does not require the $T \to 0$ limit and can thus 
be applied in any asymptotically AdS metric.}

\begin{figure}[t]
  \centering
  \includegraphics[width=.8\textwidth]{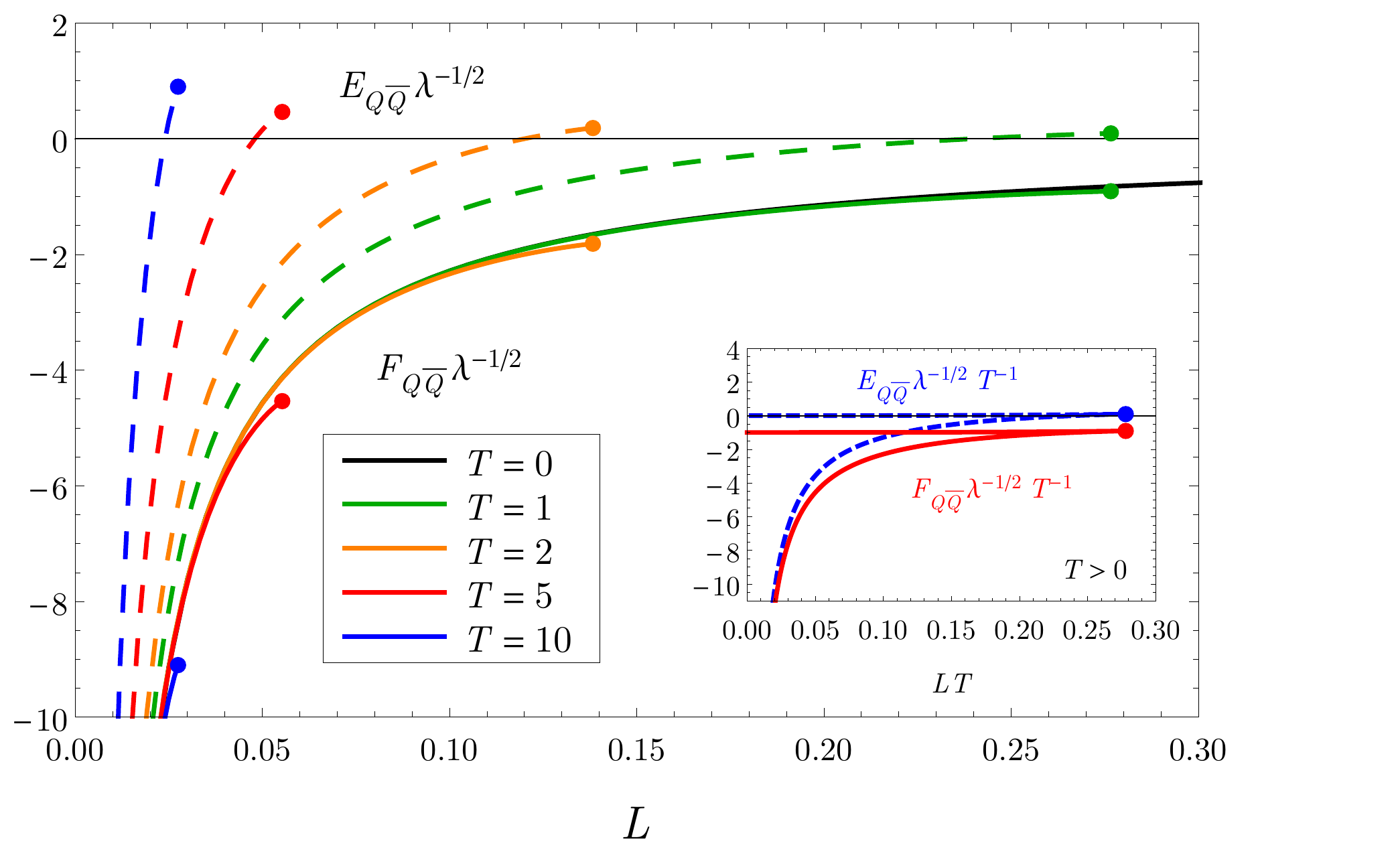}
  \caption{%
    Free energy (solid curves) and binding energy (dashed curves) in
    $\mathcal{N}=4$ SYM for varying temperature $T$, restricted to
    the stable branch (main plot) and including both the stable and
    the unstable branch (inset, $T>0$), see text. %
    For $T=0$, both $\FQQ$ and $\EQQ$ reduce to the same Coulombic
    potential (solid black curve) given by \eqref{eq:77}. %
    In the main plot, we express all dimensionful quantities in AdS
    units specified by $\LAdS=1$, and in the inset in units of
    temperature. %
    The dots on the endpoints of the curves mark the screening
    distance.%
    \label{fig:E_vs_B_N4_varT}
  }
\end{figure}

Let us now consider again general temperature $T$. 
In fig.~\ref{fig:E_vs_B_N4_varT}, we plot $\FQQ(L)$ and $\EQQ(L)$ in
$\mathcal{N}=4$ SYM for varying temperature.
We recall that we calculate these observables only for distances 
up to the screening distance $\Ls$ which we indicate here and in 
the following figures by dots at the ends of the respective curves. 
Both $\FQQ(L)$ and $\EQQ(L)$ actually have two values for every
distance $L$ smaller than the screening distance $\Ls$, \ie, both
functions have two branches. This is a consequence of there being 
two string configurations for every distance $L<\Ls$, as discussed 
in sec.~\ref{sec:string-gener-metr}. 
The inset in fig.~\ref{fig:E_vs_B_N4_varT}
displays the full $\FQQ(L)$ and $\EQQ(L)$, showing their lower and
upper branches.
The lower branches correspond to the string configurations that stay
closer to the boundary. Since their free energy is smaller than that
of the string configurations protruding farther into the bulk, they
are energetically preferred.
In addition, it turns out that the solutions that reach farther
into the bulk possess runaway modes when subjected to small
perturbations whereas the string configurations that stay closer to
the boundary are stable against such perturbations
\cite{Friess:2006rk}.
From now on, we will always restrict our discussion to the stable, lower branches of
both $\FQQ$ and $\EQQ$, and accordingly we have not plotted the upper
branches in the main plot in fig.~\ref{fig:E_vs_B_N4_varT}. 

In fig.~\ref{fig:E_vs_B_N4_varT} we observe several characteristic 
properties of the free energy and the binding energy. 
For any value of $T$, the free energy $\FQQ$ becomes Coulombic
at small inter-quark distances. This signals a restoration of
conformality in the UV as the medium-induced scale $T$
decouples. We further observe that the free energy becomes 
independent of $T$ for small $L$. 
The binding energy $\EQQ$, on the other hand, depends on $T$ 
also for small distances $L$ as it contains
an $L$-independent but $T$-dependent contribution. 
Fig.~\ref{fig:E_vs_B_N4_varT} also shows that the two energies 
exhibit a qualitatively different dependence on temperature. 
For $T=0$, both
$\FQQ(L)$ and $\EQQ(L)$ reduce to the same function, namely the
zero-temperature potential given in \eqref{eq:77}. For
increasing $T$, the free energy at fixed distance $L$ becomes smaller 
as compared to the $T=0$ limit, as is expected due to the screening 
of color charges in the medium. In contrast to that, 
$\EQQ$ increases with temperature.
This behavior of $\EQQ$ is consistent with our interpretation of it as
a (negative) binding energy. Due to increasing $T$, at a fixed
distance $L$ the modulus of the binding energy decreases. In other words,
the binding of the quarks becomes weaker, as is natural 
in a hotter medium due to stronger screening of the interaction.
Accordingly, also the screening distance becomes smaller for
increasing $T$.

For any non-zero temperature, at some value
$\Lc<\Ls$ of the inter-quark separation, the binding energy vanishes,
$\EQQ(\Lc)=0$. Thus, at this distance the free energy of the bound
$\QbarQ$ pair equals the free energy of an unbound $\QbarQ$ pair,
while for larger distances the free energy of an unbound pair is
smaller than that of a bound pair. However, this does not necessarily
imply that the $\QbarQ$ pair dissociates at this length scale. In
fact, the dynamic evolution from a bound state 
to two separate quarks would in our approach presumably 
involve more complicated string configurations similar to those 
relevant at distances $L> \Ls$, see the discussion in 
sec.\ \ref{sec:string-gener-metr}. Although the relevant string 
configurations are not known in detail, we expect the transition 
from the simple string to these configurations to include 
string breaking effects. The dynamical evolution of $\QbarQ$ 
dissociation is therefore beyond evaluating the 
approximation \eqref{eq:34} for the simple string configuration 
discussed above. 
The $\QbarQ$ pair might well be metastable even beyond
$\Lc$. For further discussion of this issue see, \eg,
\cite{Friess:2006rk}. 

At this point we make an interesting observation. 
In contrast to the binding energy $\EQQ$, the free energy $\FQQ$ 
of the pair does not exhibit a zero at any distance below $\Ls$, 
see fig.~\ref{fig:E_vs_B_N4_varT}. 
(In that figure we have fixed the overall constant (\ie{}, the renormalization 
scheme) such that $\FQQ$ coincides with the zero-temperature 
heavy-quark potential at small $L$, see eq.\ \eqref{eq:77}. 
With a different subtraction scheme, a zero could be made occur at 
an arbitrary point (below $\Ls$) and thus cannot have any physical meaning.) 
It is worth pointing out that the considerations made in \cite{Bak:2007fk} 
concerning the behavior of $\FQQ$ just above $\Ls$ are based on 
a general convexity property of the free energy and make use of 
a zero of the putative quark--antiquark free 
energy (for which $\EQQ$ was taken there as a consequence of not 
using the correct renormalization). In absence of such a zero in the 
actual free energy $\FQQ$, the arguments for the smallness of the 
transition region between different dominant string configurations 
around $\Ls$ need to be reconsidered, as we have mentioned 
already in sec.~\ref{sec:string-gener-metr}.

\begin{figure}[t]
  \centering
  \includegraphics[width=.8\textwidth]{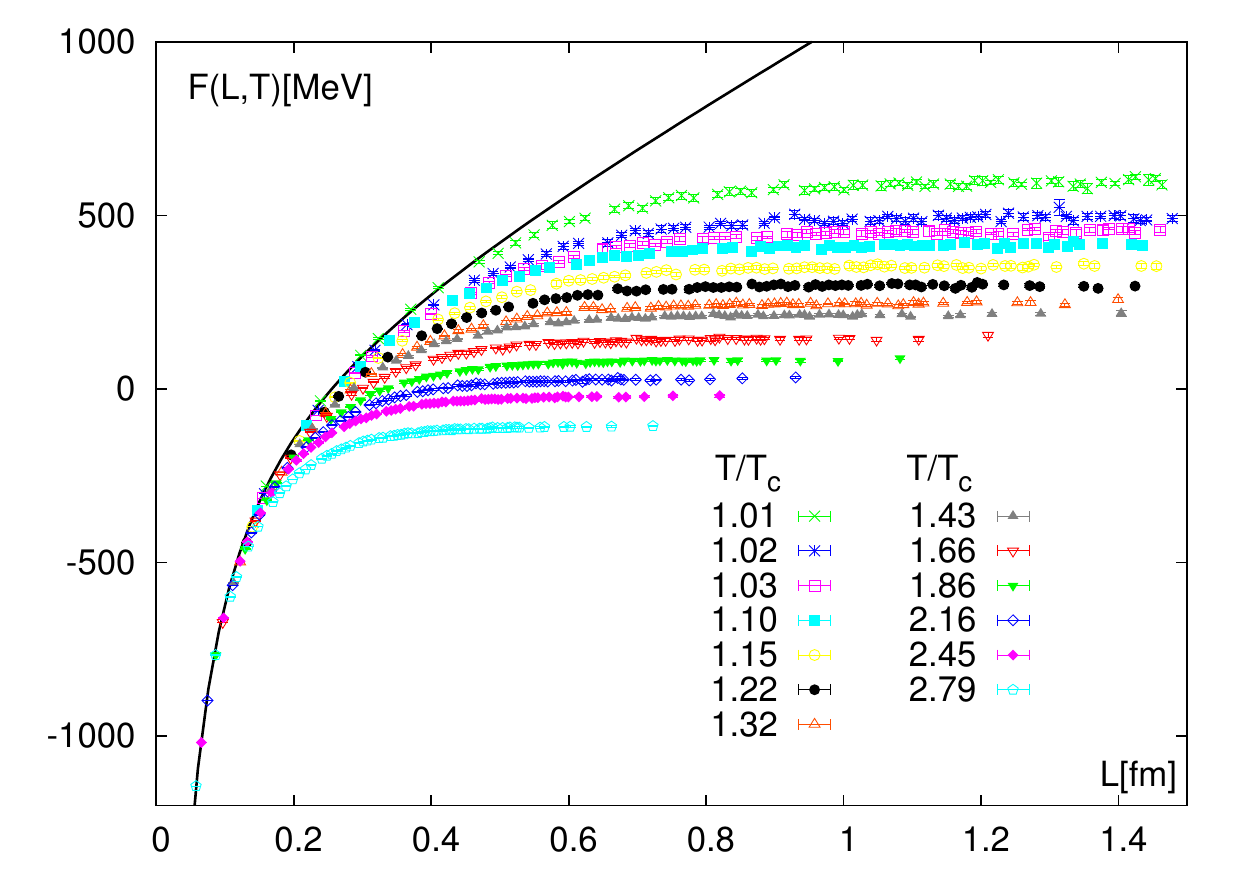}
  \caption{%
    Heavy-quark free energy at various temperatures above $T_c$ 
    from a 2+1-flavor lattice QCD calculation \cite{Kaczmarek:2007pb}. 
    The solid line shows the zero temperature heavy-quark potential \cite{Cheng:2007jq}. 
    \label{fig:FQQ_quenched-lattice}
  }
\end{figure}

Next, we want to compare the qualitative behavior of
the free energy in $\mathcal{N}=4$ SYM to that of QCD. We clearly do not 
expect an exact quantitative agreement of $\FQQ$ in these two theories. 
However, the qualitative effect of the thermal medium on the heavy quark pair 
should be largely independent of the microscopic degrees of freedom 
present in the two theories. In lattice QCD, the
heavy-quark free energy can be extracted from a correlator of Polyakov
loops.
In fig.~\ref{fig:FQQ_quenched-lattice} we show lattice QCD results 
from \cite{Kaczmarek:2007pb} for 2+1-flavors with a physical strange 
quark mass and a pion mass of around 220~MeV using an improved staggered 
quark action \cite{Cheng:2007jq}. For a discussion of the renormalization 
of the heavy quark free energies in lattice QCD see \cite{Gupta:2007ax}. 
The temperature is varied and all chosen temperatures are above
$\Tc$. We note two characteristics of the behavior of the free
energy. First, for small inter-quark distances $L$ 
the free energy becomes independent of the
temperature. Second, the free energy decreases with increasing $T$,
\ie, data points for some $T_2>T_1$ always lie below those for $T_1$.
Both these characteristics are also present in the free energy
computed in $\mathcal{N}=4$ SYM, as discussed above. Furthermore, they
are also present in our non-conformal models that will be discussed in
the next section.
In contrast to this, the quantity $\EQQ$ behaves differently. It
rather increases with increasing $T$.
These findings further substantiate our general arguments regarding
the choice \eqref{eq:50} of the subtraction $\Delta S$. Using it in
\eqref{eq:47} we indeed obtain the proper $\QbarQ$ free energy.
We again recall that in the holographic calculation we consider 
only interquark distances up to the temperature-dependent screening 
distance $\Ls$, see sec.\ \ref{sec:string-gener-metr}. 
We therefore cannot expect to reproduce in these calculations 
the full transition to a flattening potential at large distances 
exhibited in the QCD lattice data in fig.\ \ref{fig:FQQ_quenched-lattice}. 
At the distances that we can address, the resulting free energy $\FQQ$ 
shows the correct behavior. 

\section{Free energy and binding energy in non-conformal models}
\label{sec:invest-non-conf}

Next, we want to study the free energy and the binding energy of a heavy 
quark--antiquark pair in non-conformal theories, that is in theories in 
which the conformal symmetry is broken explicitly and not only by 
temperature. One can obtain such 
non-conformal theories as duals of suitable deformations of the 
AdS-Schwarzschild metric, typically containing one or more dimensionful 
parameters. Various non-conformal metrics have been discussed in 
the literature. Many of them aim at reproducing 
properties of the actual QCD plasma. 
Specific properties of QCD and their implementation in holographic 
models are expected to become particularly relevant in the infrared, that is 
at large distances. 
As explained in sec.\ \ref{sec:string-gener-metr}, in our present study 
we calculate the finite-temperature observables only for relatively small distances, namely 
up to the respective screening distance $\Ls$. These observables are 
therefore not very sensitive to the infrared properties of QCD-like 
holographic models. We therefore take a different 
approach and look for generic properties or universal behavior 
emerging in large classes of non-conformal bottom-up models. 
To this end, we consider some prototype models which have been 
used before to study various observables in this spirit. 
In the first family of models (\SWT{} model) the non-conformality is 
introduced by hand, in the second set of models 
(consistently deformed 1-parameter models) it is associated 
with additional scalar fields in the bulk. In the latter models the metric 
solves the equations of motion for a five-dimensional 
Einstein--Hilbert--scalar action. 
We will not assume that these models can be embedded into 
a higher-dimensional string theory. 
We will briefly present the main properties of 
the models that we consider before we proceed to discussing the heavy quark pair 
in these backgrounds. 

\paragraph{\SWT{} model.}
This model has been introduced in \cite{Andreev:2006eh,Kajantie:2006hv} 
and is motivated by the soft-wall model at zero temperature \cite{Karch:2006pv} 
which had considerable success in describing various aspects of low-energy 
hadron physics. In the soft-wall model the AdS metric corresponding to 
$\mathcal{N}=4$ SYM at zero temperature is multiplied by an overall warp 
factor of the form $\e^{c^2z^2}$ with some deformation parameter $c$ 
that determines the deviation from conformality. 
Applying the same procedure to the AdS-Schwarzschild metric corresponding to 
$\mathcal{N}=4$ SYM at finite temperature one obtains the 
one-parameter family of models \cite{Andreev:2006eh,Kajantie:2006hv} 
\begin{equation}
  \label{eq:SWT}
  \begin{aligned}
    \d s^2 &= \frac{\LAdS^2 \, \e^{c^2 z^2}}{z^2} \left(-h(z) \d t^2 + \d\vec{x}^2
      + \frac{1}{h(z)} \d z^2\right) \,,\\
    h(z) &= 1 - \frac{z^4}{\zh^4} \,,
  \end{aligned}
\end{equation}
which we call the \SWT{} models, for `soft wall-like
models at finite temperature $T$' \cite{Schade:2012ah}. 
The horizon function $h$ is the same as in the AdS-Schwarzschild 
metric \eqref{eq:78}, to which the metric \eqref{eq:SWT} reduces in the 
limit $c\to 0$. Accordingly, the temperature of the boundary field theory 
in these models is defined as in $\mathcal{N}=4$ SYM, 
\begin{equation}
T=\frac{1}{\pi \zh} \,.
\end{equation}
Various observables concerning heavy quarks in the medium have 
been studied in the \SWT{} model, see for example 
\cite{Andreev:2006eh,Andreev:2006nw,Kajantie:2006hv,Nakano:2006js,Liu:2008tz}. 
A suitable choice for the parameter $c$ of the model can be made 
by comparing observables to their phenomenological values in the 
actual quark--gluon plasma. In \cite{Liu:2008tz} it is argued that 
$0 \le c/T \le 2.5$ is a reasonable range for that choice.\footnote{Note that 
in our warp factor, $\exp(c^2z^2)$, the deformation parameter enters quadratically 
while \cite{Andreev:2006eh} or \cite{Kajantie:2006hv,Liu:2008tz}  
have $\exp(cz^2)$ or $\exp\left(\frac{29}{20} cz^2 \right)$, respectively. 
This is just a different notation, and it is straightforward to convert 
the corresponding values of the deformation parameter. The values 
quoted in the present paper apply to our choice of warp factor. We also 
recall that we are interested here in the general behavior under 
non-conformal deformations rather than in determining a precise 
value for the deformation.} 
The advantage of the \SWT{} model is its simplicity which even permits 
to compute some observables analytically. However, the metric \eqref{eq:SWT} 
does not solve the equations of motion of any five-dimensional gravity 
action and is in that sense not consistent. As a consequence, the 
corresponding boundary theory does not satisfy general thermodynamic 
relations \cite{Kajantie:2006hv}. 
This problem is solved, though at the expense of a higher computational 
effort, by consistently deformed models to which we turn next. 
Despite its limitations, the \SWT{} model can be useful as a simple method to obtain 
a first impression of the behavior of many observables 
under non-conformal deformations, as will also be confirmed by our results below. 

\paragraph{Consistently deformed 1-parameter models.}
In general, it is preferable to study metrics that solve the Einstein 
equations of a gravity action, as the dual theories of such models 
are expected to exhibit consistent thermodynamic observables. 
One can construct non-conformal metrics 
of this kind by considering a bulk scalar field in AdS and its backreaction 
on the AdS space \cite{Gubser:2008ny}. 
This setup is described by the five-dimensional Einstein--Hilbert--scalar action 
\begin{equation}
  \label{eq:action}
  S = \frac{1}{16\pi\GN^{(5)}} \int\d^5x\, \sqrt{-g} 
\left( \mathcal{R} - \frac{1}{2} \partial_M\phi\, \partial^M\phi - V(\phi) \right) \,,
\end{equation}
where $g$ is the determinant of the metric $g_{MN}$, $\mathcal{R}$
is the associated Ricci scalar, and $\GN^{(5)}$ is the
five-dimensional Newton constant. $\phi$ is the scalar that will 
induce the deformation away from conformality. 
The potential $V(\phi)$ for the scalar is assumed to contain as a 
constant term 
\begin{equation}
  \label{eq:Vphi0}
V(\phi=0) = - \frac{12}{\LAdS^2}
\end{equation}
which is twice the cosmological constant of an AdS space 
with curvature radius $\LAdS$. For a vanishing scalar $\phi$ 
one obtains from \eqref{eq:action} a pure AdS space \eqref{eq:78}
corresponding to a conformal boundary theory, 
and then only the temperature $T$ is an additional parameter of the 
metric. 
Starting from the ansatz \eqref{eq:metric} for the general metric 
one can derive a coupled set 
of differential equations for the functions $A$, $B$, $h$, and $V$ 
from the equations of motion of the action \eqref{eq:action} 
\cite{Gubser:2008ny}. One option is to start by specifying a 
potential $V$, see for example \cite{Gubser:2008ny}. 
The coupled equations also permit another possibility, namely 
to specify the scalar profile in $z$ and to calculate a suitable 
potential $V$ \cite{DeWolfe:2009vs}. 
We choose the latter option as it allows us to study models 
that come very close to the phenomenologically well-motivated 
soft-wall model. Also in this case it is possible to study large 
classes of metrics with one or more parameters. 
A two-parameter model of this kind has been proposed in \cite{DeWolfe:2009vs}. 
In \cite{Schade:2012ah} a simplified 1-parameter version 
of that model has been devised which has the advantage that all 
functions in the metric can be expressed in closed form. We will 
in the following work with this 1-parameter model as it captures 
the main features relevant for our considerations. 
Before we describe the model in detail, a comment is in 
order concerning its general construction. We fix the same scalar 
profile in $z$ and then for each temperature calculate the scalar 
potential $V(\phi)$ from the equations of motion of \eqref{eq:action}. 
In general, this leads to different potentials $V(\phi)$ for 
different temperatures, which implies that we consider 
different theories for different temperatures. This is, strictly speaking, 
an inconsistent procedure. However, it turns out that for the 
specific models that we consider here this approach is 
acceptable from a practical perspective, as was shown for 
the 2-parameter model in \cite{DeWolfe:2009vs} and for the 
1-parameter model in \cite{Schade:2012ah}. Our choice of 
scalar potential will be quadratic in $z$ as in the soft-wall 
model, and therefore large values of the scalar $\phi$ 
correspond to large $z$. Solving for $V(\phi)$ one finds that 
the potentials for different temperatures follow a universal 
curve up to values of $\phi$ corresponding to the respective 
horizon position for the chosen temperature. Up to the respective 
horizon, the deviation from the universal curve is numerically very 
small. Our observables are computed from string configurations 
for which only the region above the horizon is relevant, and 
the differences in the scalar potentials at different temperatures 
have a negligible effect on these strings. 
Therefore, we here follow the practical approach 
to fix the scalar potential as this method is numerically 
simpler and allows a better comparison to the \SWT{} model. 

To obtain the one-parameter  
model, one makes in the general metric \eqref{eq:metric}
the ansatz 
\begin{equation}
  \label{eq:1Pansatz}
  \begin{aligned}
    \phi(z) &= \sqrt{\frac{3}{2}}  \kappa z^2 \,, \\
    A(z) &= \log\left(\frac{\LAdS}{z}\right) 
  \end{aligned}
\end{equation}
with a dimensionful deformation parameter $\kappa \ge 0$. 
The equations of motion of \eqref{eq:action} then lead to 
\begin{equation}
  \label{eq:1PB}
  B(z) = \log\left(\frac{\LAdS}{z}\right) - \frac{1}{4} \,\kappa^2 z^4 \,. 
\end{equation}
Also $h$ and $V$ can be obtained in closed form. We do not give these 
somewhat lengthy expressions as their details will not be relevant for the 
following discussion. With these functions one obtains from \eqref{eq:91} 
the temperature $T$ in terms of the horizon position $\zh$ in the 
1-parameter model, 
\begin{equation}
  \label{eq:1}
  T = \frac{1}{\pi\zh} \frac{\kappa^2\zh^4}{4} 
  \frac{\e^{\kappa^2\zh^4/4}}{\e^{\kappa^2\zh^4/4} - 1}
  \,.
\end{equation}
We will in the following consider typical values of the deformation 
parameter\footnote{In 
the 1-parameter model it turns out that for a given temperature 
there is in fact a maximal deformation $\kappa$ beyond which no 
solution can be found for that temperature \cite{Schade:2012ah}, given by 
$(\sqrt{\kappa} / T)|_{\text{max}} = 2.94$.
Our choice of the deformation parameter $0 \le \sqrt{\kappa}/T \le 2.5$ 
covers almost all of the possible range.\label{foot4}}
in a similar range as in the \SWT{} model, $0 \le \sqrt{\kappa}/T \le 2.5$. 
In the limit $\kappa \to 0$ the 1-parameter model reduces to the 
pure AdS metric, and hence to a conformal boundary theory. 
Finally, the scalar $\phi$ in the 1-parameter model can be but need 
not be the dilaton. 
In the bottom-up models that we consider here,  
this is simply a choice one can make, and this alternative gives in fact 
rise to two distinct versions of the 1-parameter model, \cf\ a similar 
discussion in \cite{DeWolfe:2009vs}. 
(This choice would no longer exist if the model were derived 
from a higher-dimensional string theory in which case the dilaton 
would be distinguished from other scalars. In our bottom-up models 
we assume no such embedding into a string theory.) 
The difference between the 
two versions lies in the metric used in the calculation of the 
string configuration as outlined in sec.~\ref{sec:string-gener-metr}. 
The metric \eqref{eq:metric} with the functions $A$, $B$, and $h$ 
just described is called the Einstein-frame metric, $g^{(\text{E})}$. 
The string configuration has to be calculated using the string-frame metric, 
$g^{(\text{s})}$, which is obtained by multiplying the Einstein-frame metric 
by a factor containing the dilaton, 
\begin{equation}
  \label{eq:frametransition}
  g^{(\text{s})} = \e^{\sqrt{\frac{2}{3}} \phi_\text{dilaton}} \, g^{(\text{E})} \,.
\end{equation}
If $\phi$ is the dilaton, the string-frame metric hence differs from the 
Einstein-frame metric by the warp factor in \eqref{eq:frametransition} 
with the scalar profile in \eqref{eq:1Pansatz}. In a model without 
a dilaton, \ie\ if $\phi$ is some other scalar, the two frames coincide. 
To distinguish the models, we therefore 
speak of the `string frame' version of the 1-parameter model 
if $\phi$ is the dilaton, and of the `Einstein frame' version if $\phi$ is 
not the dilaton. The additional warp factor of the string-frame 
model makes this version of the 1-parameter metric rather similar to the 
\SWT{} metric, and the qualitative similarity of these two models will 
also be seen in our results below. 

\begin{figure}[t]
 \centering
 \includegraphics[width=.8\textwidth]{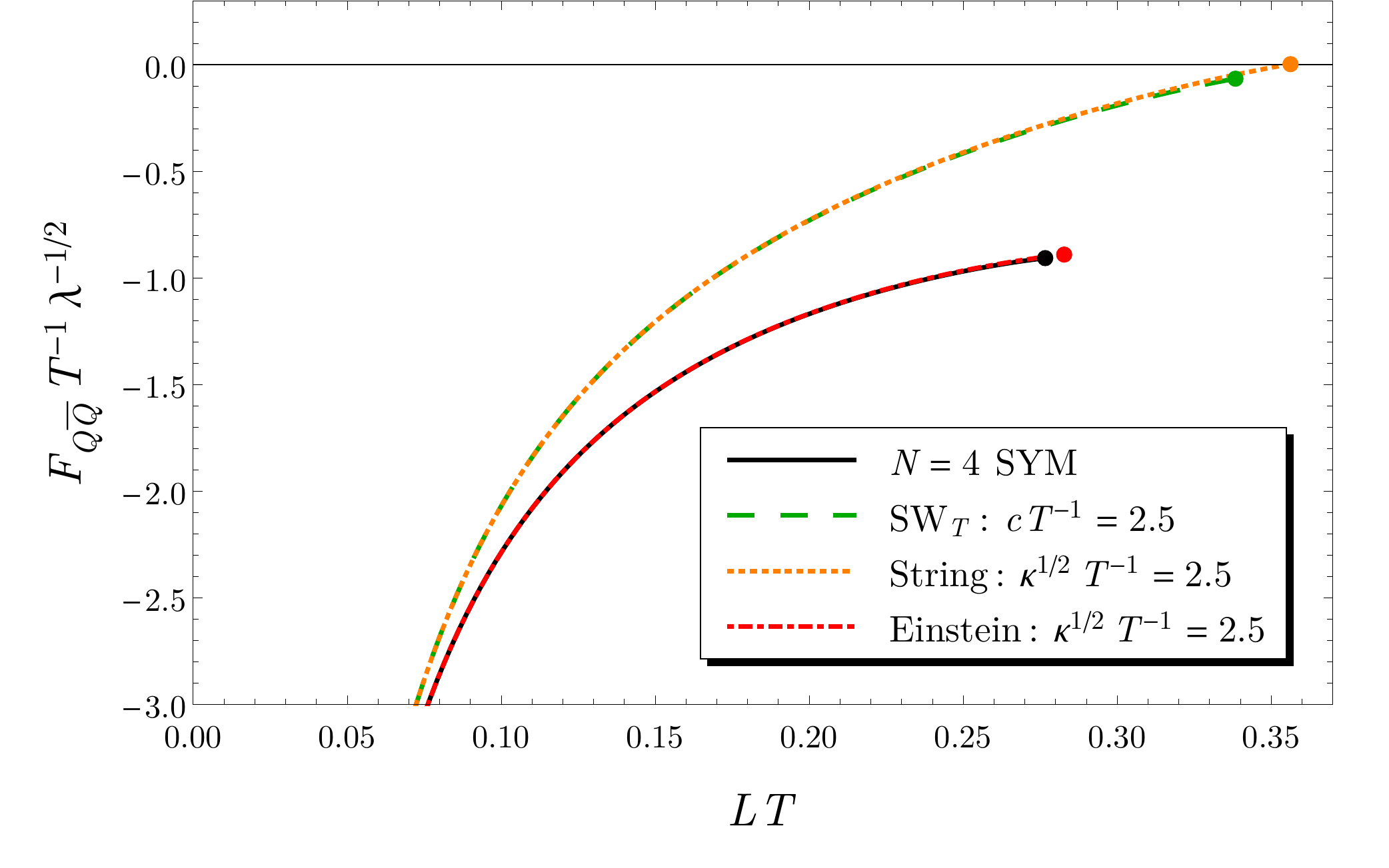}
 \caption{%
   Free energy $\FQQ$ in $\mathcal{N}=4$ SYM and in
   the non-conformal models at large dimensionless
   deformation-parameter-to-temperature ratios, $c/T$ and
   $\sqrt\kappa/T$, respectively, for fixed temperature. %
   To be able to discern details of the free energy close to the
   screening distance (marked by dots), we do not show the curves for
   very small distances. For small $LT$, all shown curves converge to
   one universal curve. All dimensionful quantities are measured in
   units of the temperature.%
   \label{fig:F_all_models_largeDef}
 }
\end{figure}

\paragraph{}
Let us now turn to the free energy of the heavy quark--antiquark pair 
in these non-conformal models. In fig.~\ref{fig:F_all_models_largeDef} 
we show the free energy $\FQQ$ at a fixed temperature and a large value
of the dimensionless ratios of the deformation parameters 
and the temperature, $c/T =2.5$ and $\sqrt\kappa/T=2.5$ 
for the \SWT{} model and the 1-parameter models, respectively.
For comparison, we have also plotted $\FQQ$ in $\mathcal{N}=4$ SYM (black
curve). Note that here and in some of the plots below 
we show all quantities in units of temperature, 
that is we use dimensionless ratios on the axes, here $\FQQ/T$ and $LT$. 
For non-conformal theories, however, the values 
of the observables for a different temperature $T'$ cannot be read off 
from the same curve, instead one would have to regard the 
curve corresponding to an accordingly changed ratio of deformation 
parameter and temperature, $c/T'$ or $\sqrt{\kappa}/T'$. 
We show such different curves only in some cases where we vary 
the deformation parameter.
In spite of this slight complication in the interpretation of the 
curves, the representation in terms of dimensionless quantities 
appears to be the best way to compare different theories, that is 
models with different non-conformal deformation parameters $c$ or $\kappa$.

For quark separations $L$ somewhat smaller than those
shown in fig.~\ref{fig:F_all_models_largeDef}, which we have
left out of the plot to not obscure the details close to the screening
distance (marked by the dots), the free energy approaches a single
universal curve even for different models. That curve is
given by the vacuum potential $\VQQ$ of $\mathcal{N}=4$ SYM, see
\eqref{eq:77}. This confirms our considerations concerning the 
small-distance behavior of $\FQQ$ in terms of the bulk picture 
in sec.~\ref{sec:general-discussion}.

Since the curves for all our models approach a single universal curve
for small quark separation, we can sensibly compare the free energy in
different theories.%
\footnote{\textit{A priori}, the free energy is only defined up to an
  overall constant offset. However, by demanding that the free energy
  approaches the vacuum potential $\VQQ$ for small distances this
  ambiguity is fixed.}
In particular, let us compare the different non-conformal models to
$\mathcal{N}=4$ SYM. First we note that the 1-parameter
Einstein-frame model is very robust against non-conformal
deformation. The free energy in this model is only slightly above 
its value in $\mathcal{N}=4$ SYM, even at the relatively large deformation 
considered here. 
On the other hand, in both the \SWT{} and 1-parameter string-frame
models the free energy increases well above its value in $\mathcal{N}=4$
SYM upon introducing non-conformality. 
As a common property we observe that in all our models the 
free energy in the non-conformal models is above that in $\mathcal{N}=4$ SYM. 
Hence $\FQQ$ in $\mathcal{N}=4$ SYM
appears to be a lower bound for estimating the free energy
of a heavy quark--antiquark pair, if the latter is normalized such
that for small distances it reduces to the potential at $T=0$ given in
\eqref{eq:77}. We expect this bound to apply to a large class 
of non-conformal holographic theories. 

\begin{figure}[t]
  \centering
  \includegraphics[width=.8\textwidth]{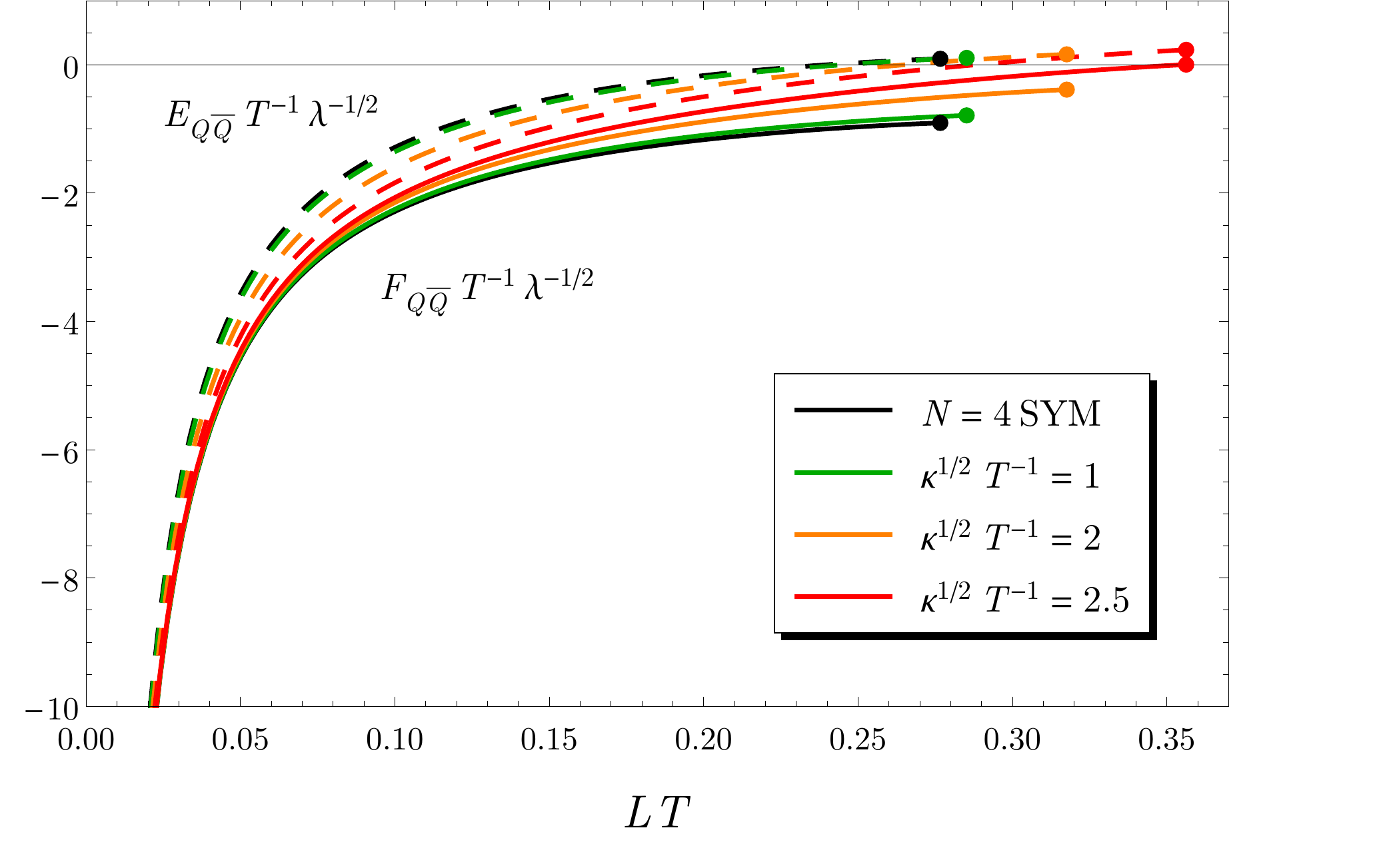}
  \caption{%
    Free energy $\FQQ$ (solid curves) and binding energy $\EQQ$
    (dashed curves) in $\mathcal{N}=4$ SYM and the
    1-parameter string-frame model at fixed temperature for varying
    deformation parameter. All dimensionful quantities are measured in
    units of the temperature.%
    \label{fig:E_vs_F_1pStr_VarKappa}
  }
\end{figure}

Next, we compare the free energy $\FQQ$ and the binding
energy $\EQQ$ in the 1-parameter string-frame model, as an example of a
consistent non-conformal deformation of $\mathcal{N}=4$ SYM. This will
also allow us to check the statement we just made about the
lower bound for the free energy at smaller values of the
deformation.
In fig.~\ref{fig:E_vs_F_1pStr_VarKappa} we plot $\FQQ$ and $\EQQ$ in
the 1-parameter string-frame model for
varying dimensionless deformation parameter $\sqrt\kappa/T$.
We see that in this model the free energy
gradually increases with increasing deformation. Thus, it indeed
always stays above its value in $\mathcal{N}=4$ SYM, even for smaller
deformations. The figure also shows that, as mentioned above, the
free energy in different theories (corresponding to different values
of the deformation) converges upon a single universal curve for small
quark separations $L$.

We further observe in fig.~\ref{fig:E_vs_F_1pStr_VarKappa} that, 
in contrast to the behavior of the free energy, the binding energy 
$\EQQ$ decreases with
increasing deformation. Thus, the binding of the $\QbarQ$ pair at a
given $L$ becomes stronger. (Recall that $\EQQ$ is actually the
negative binding energy.) 
We have studied the behavior of the binding energy with respect to the
deformation also in our other non-conformal models. 
We find that, as was the case for the free energy, the
binding energy $\EQQ$, too, behaves very similarly in the \SWT{} model
and in the 1-parameter string-frame model for which we have shown
$\EQQ$ in fig.~\ref{fig:E_vs_F_1pStr_VarKappa}. Furthermore, this
quantity is very robust in the 1-parameter Einstein-frame model,
staying quantitatively close to its counterpart in $\mathcal{N}=4$ SYM
for all values of the deformation parameter.
Also the feature that clearly distinguishes the two quantities $\FQQ$
and $\EQQ$ from each other turns out to be universal in all our 
non-conformal models: while $\FQQ(L)$ at fixed $L$ increases with
increasing non-conformality, $\EQQ(L)$ decreases with increasing 
non-conformality. $\EQQ(L)$ in $\mathcal{N}=4$ SYM 
might possibly be an upper bound for the binding energy of a heavy 
quark--antiquark pair in a large class of non-conformal theories. 

\section{Entropy and internal energy of a heavy quark pair}
\label{sec:heavy-quark-entropy}

In this section we want to discuss two quantities that can be derived 
from the free energy $\FQQ$ of the heavy quark--antiquark pair, 
namely the entropy and the internal energy of the pair. For the derivation of these 
observables it is crucial to use the correct (temperature-independent) 
subtraction in the renormalization \eqref{eq:47} of the free energy. 

The entropy and the internal energy of the heavy quark--antiquark pair 
can be computed from the free energy using standard thermodynamic 
relations. In a given holographic model the heavy-quark free energy 
$\FQQ(L,T)$ depends on the inter-quark distance $L$ and the 
temperature $T$ of the medium. The entropy $\SQQ$ of the pair 
can be computed (following the definition also used in lattice QCD, 
see for example \cite{Kaczmarek:2005gi}) as the derivative 
\begin{equation}
  \label{eq:67}
  \SQQ(L,T) = -\frac{\del\FQQ(L,T)}{\del T} \,.
\end{equation}
With the entropy in hand, the internal energy can be obtained from 
\begin{equation}
  \label{eq:68}
  \UQQ(L,T) = \FQQ(L,T) + T\SQQ(L,T) \,.
\end{equation}
The explicit computation in our holographic models, where we have the
parametric expressions \eqref{eq:51} for $\FQQ$ and \eqref{eq:35} for
the distance $L$ in terms of the bulk length scales $\zt$ and $\zh$,
is not entirely straightforward. We give details on the computation
and an explicit formula for the derivative $\del\FQQ/\del T$ in
appendix~\ref{sec:deta-comp-qbarq}.

Both the free and the internal energy are phenomenologically
interesting as candidates for model potentials for the interaction of
heavy quarks in a finite-temperature medium. Model potentials are used
for the computation of properties of heavy quarkonia from
Schr\"odinger-like equations in the spirit of potential
non-relativistic QCD (pNRQCD; see \cite{Brambilla:2004jw} for a
review, and \eg~\cite{Brambilla:2008cx} for more recent work including
finite-temperature effects).  At zero temperature, pNRQCD provides a
systematic framework for the derivation of an effective $\QbarQ$
potential. At non-zero temperature the choice of a potential to model
the heavy-quark interaction is to some extent ambiguous.
The internal and the free energy differ from each other due to the
entropy contribution, and it is thus worth exploring the behavior of
both of these quantities. See also, for instance,
\cite{Petreczky:2005bd,Kaczmarek:2005gi,Burnier:2015tda} for
discussions of heavy-quark energies and potentials in the context of
lattice QCD.

Let us first compute the $\QbarQ$ entropy and
internal energy in $\mathcal{N}=4$ SYM. The
above formulae \eqref{eq:67} and \eqref{eq:68} can be evaluated
explicitly based on the expressions \eqref{eq:61} and \eqref{eq:62}
for $\FQQ(\zt)$ and $L(\zt)$, plugged into \eqref{eq:66} in the
appendix. We obtain the parametric expressions 
\begin{equation}
  \label{eq:69}
  \frac{\SQQ(\zt)}{\sqrt\lambda}
  = \frac{2\sqrt\pi\Gammafct{\frac{3}{4}}}{\Gammafct{\frac{1}{4}}}\,
  \frac{\zh}{\zt}\,
  \frac{5\left(1-\frac{\zt^4}{\zh^4}\right)\fa^2
  +\left[3\left(1-\frac{\zt^4}{\zh^4}\right)\fb-5\fa\right]\fc}%
  {5\left(\frac{\zh^4}{\zt^4}-3\right)\fa+6\left(1-\frac{\zt^4}{\zh^4}\right)\fb}\,,
\end{equation}
and
\begin{equation}
  \label{eq:71}
  \frac{\UQQ(\zt)}{\sqrt\lambda}
  = -\frac{5\Gammafct{\frac{3}{4}}}{\sqrt\pi\Gammafct{\frac{1}{4}}}\,
  \frac{1}{\zt}\,
  \frac{\left(1-\frac{\zt^4}{\zh^4}\right)\fa\fd}%
  {5\left(1-3\frac{\zt^4}{\zh^4}\right)\fa
    + 6\frac{\zt^4}{\zh^4}\left(1-\frac{\zt^4}{\zh^4}\right)\fb} \,,
\end{equation}
where $\fa$, $\fb$, $\fc$, and $\fd$ depend on $\zt^4/\zh^4$: we
define them as shorthand notation for the functions
\begin{align}
  \fa\left(\frac{\zt^4}{\zh^4}\right)
  &= \Fhyp{\frac{1}{2}}{\frac{3}{4}}{\frac{5}{4}}{\frac{\zt^4}{\zh^4}}
    \,,\label{eq:19}\\
  \fb\left(\frac{\zt^4}{\zh^4}\right)
  &= \Fhyp{\frac{3}{2}}{\frac{7}{4}}{\frac{9}{4}}{\frac{\zt^4}{\zh^4}}
    \,,\label{eq:94}\\
  \fc\left(\frac{\zt^4}{\zh^4}\right)
  &= \Fhyp{-\frac{1}{2}}{-\frac{1}{4}}{\frac{1}{4}}{\frac{\zt^4}{\zh^4}}
    \,,\label{eq:101}\\
  \fd\left(\frac{\zt^4}{\zh^4}\right)
  &= \Fhyp{-\frac{1}{2}}{\frac{3}{4}}{\frac{1}{4}}{\frac{\zt^4}{\zh^4}}
    \,.\label{eq:102}
\end{align}
The entropy $\SQQ$ vanishes identically in the limit $T\to 0$, as
can be verified analytically in $\mathcal{N}=4$ SYM from formula
\eqref{eq:69}. This implies that for $T=0$ the internal energy
coincides with the free energy. As we have seen before, in this case
the free energy, and thus also the internal energy, is given by the
zero-temperature potential $\VQQ$, see \eqref{eq:77}.

\begin{figure}[t]
  \centering
  \includegraphics[width=.8\textwidth]{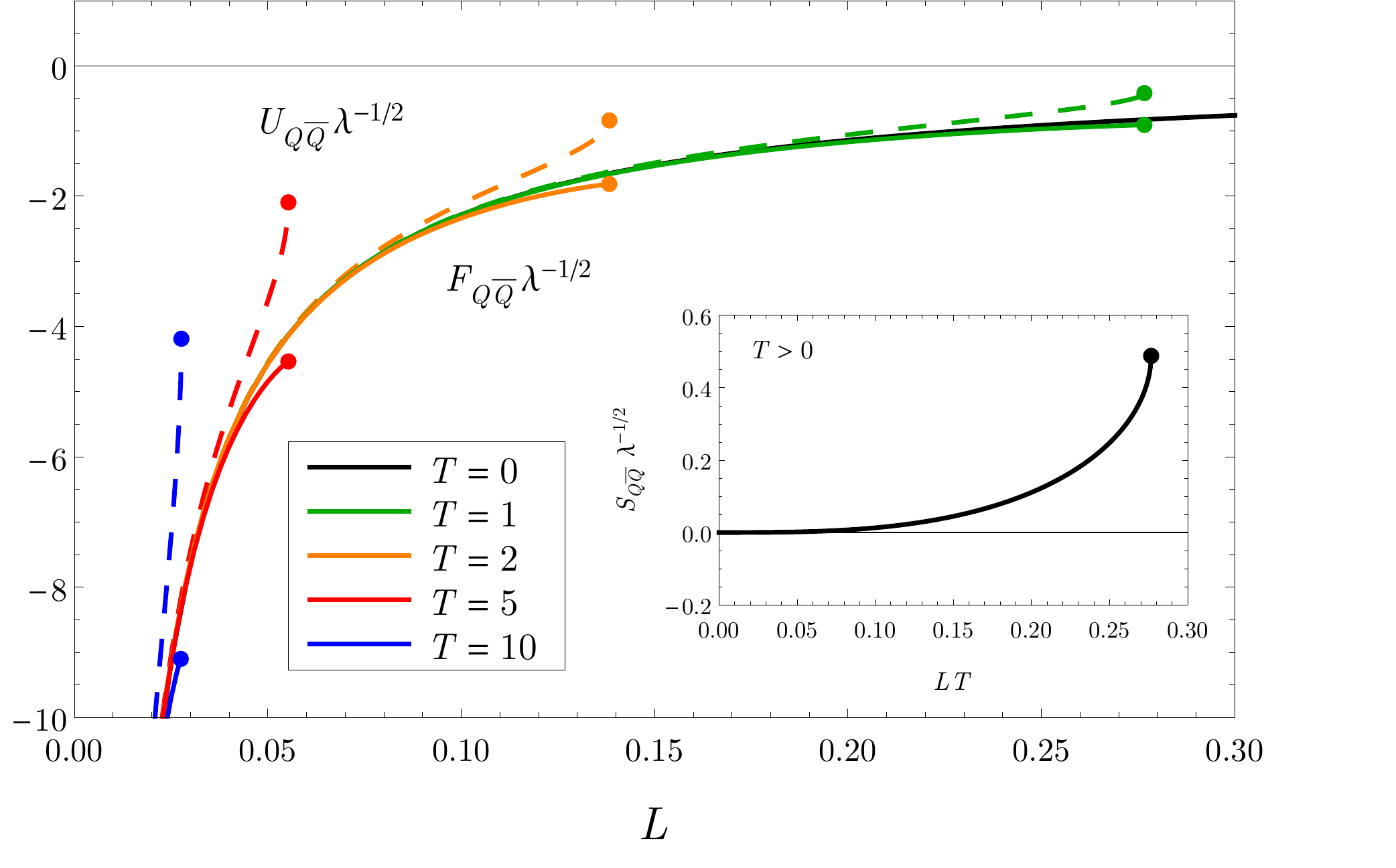}
  \caption{%
    Internal energy $\UQQ(L)/\sqrt\lambda$ (dashed curves) and free
    energy $\FQQ(L)/\sqrt\lambda$ (solid curves) for varying
    temperature in $\mathcal{N}=4$ SYM. %
    The inset shows the entropy $\SQQ/\sqrt\lambda$ for an arbitrary
    fixed $T>0$ as a function of $LT$. %
    In the main plot, we express all dimensionful quantities in
    units specified by $\LAdS=1$. %
    The dots on the endpoints of the curves mark the screening
    distance. %
    For very small $L$, the entropy approaches zero and both $\UQQ$
    and $\FQQ$ approach a universal Coulombic curve given by
    \eqref{eq:77}.%
    \label{fig:F_and_U_N4_varT}
  }
\end{figure}

In fig.~\ref{fig:F_and_U_N4_varT} we plot the
internal energy $\UQQ$ and the free energy $\FQQ$ in $\mathcal{N}=4$
SYM for increasing temperature (in AdS units set by $\LAdS=1$)
starting at $T=0$. An inset shows the entropy $\SQQ$ for any $T>0$ as
a function of the dimensionless product $LT$. Note that in
$\mathcal{N}=4$ SYM, due to the absence of any further scales, the
dimensionless entropy $\SQQ$ necessarily only depends on $LT$.
The black solid curve in fig.~\ref{fig:F_and_U_N4_varT} shows $\VQQ$.

At $T>0$, the internal and free energies start to differ from each
other and from their common $T=0$ limit $\VQQ$ at intermediate $L$ (compared
to the screening distance, which in the figure is marked by a dot on
the respective curve's endpoint).  We have discussed the behavior of
the free energy in secs.~\ref{sec:invest-mathc-supersy} and
\ref{sec:invest-non-conf}, so let us focus now on the entropy and the
internal energy.
As seen in the inset in fig.~\ref{fig:F_and_U_N4_varT}, the entropy
increases monotonically with the quark separation $L$. A heuristic
physical explanation of this observation might be that, as the size of
the $\QbarQ$ bound state increases, it has a growing overlap with the
regime of the (thermal) length scales $\Lth\sim 1/T$ of the
surrounding medium. Therefore, the $\QbarQ$ state can couple to an
increasing number of modes of the medium, thus increasing its
associated phase-space volume which leads to a rapid increase in entropy. 
With only the string configuration connecting the quark and the antiquark 
we cannot calculate the entropy of the pair for distances $L$ larger than 
the screening distance $\Ls$. When $L$ is further increased above $\Ls$ 
one expects the entropy to grow until it eventually saturates at a value corresponding 
to the entropy of two single heavy quarks, $2 S_Q$. We will calculate the latter 
in sec.~\ref{sec:single-quark-free} below. For $\mathcal{N}=4$ SYM, 
for example, a comparison of $\SQQ$ in fig.~\ref{fig:F_and_U_N4_varT} 
and $S_Q$ in fig.~\ref{fig:SQ_all_models_TOverTc} below shows that 
at $\Ls$ the entropy $\SQQ$ of the pair has reached about half the 
expected asymptotic value. 
In contrast to the free energy, the internal energy increases for
fixed $L$ upon increasing the temperature. It is always larger than
the free energy due to the positive entropy contribution $T\SQQ(L)>0$.
Interestingly, the internal energy has an inflection point and curves
upward close to the screening distance.
Since the entropy approaches zero for small distances $L$, the
internal energy approaches the free energy and shares with it the
independence of $T$ for small $L$.

\begin{figure}[t]
  \centering
  \includegraphics[width=.8\textwidth]{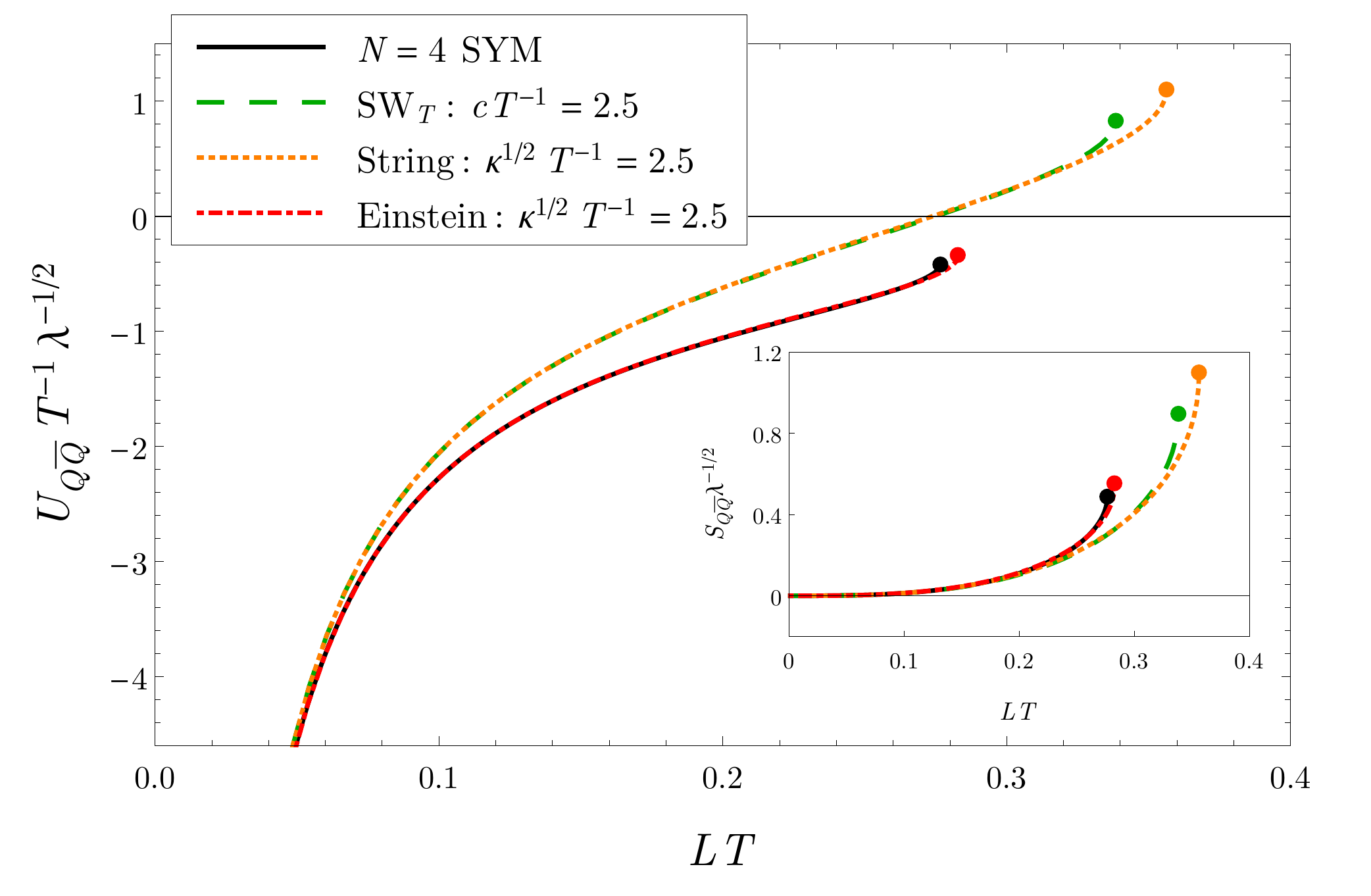}
  \caption{%
    Internal energy $\UQQ(L)/(T\sqrt\lambda)$ at fixed temperature in
    $\mathcal{N}=4$ SYM and non-conformal models at large deformations
    with $c/T=2.5$ and $=\sqrt\kappa/T=2.5$ for the
    \SWT{} and 1-parameter models, respectively. %
    The inset shows the entropy $\SQQ/\sqrt\lambda$ as a function of
    $LT$. %
    For very small $L$, the entropy approaches zero in all models and
    $\UQQ$ in all models converges to a universal Coulombic curve.%
    \label{fig:UQQ_all_models_SQQInset}
  }
\end{figure}

Having gained an understanding of the behavior of the entropy and
internal energy in $\mathcal{N}=4$ SYM, next we investigate their
behavior in our non-conformal models. Here, the entropy and 
subsequently the internal energy are computed numerically from the
free energy. The qualitative 
dependence on temperature is similar to the one in $\mathcal{N}=4$ SYM
we discussed above.
To study the impact of the deformation in more detail, in
fig.~\ref{fig:UQQ_all_models_SQQInset} we show the internal energy as
a function of the quark separation at fixed temperature in our
non-conformal models for a large value of the
dimensionless ratios of the deformation parameters and the
temperature, $c/T=2.5$ and $\sqrt\kappa/T=2.5$ in the \SWT{} and
1-parameter models, respectively.
In the inset, we plot the entropy as a function of $LT$ using the same
deformation parameters. For comparison, we also display the internal
energy and entropy in $\mathcal{N}=4$ SYM (black curves).
We see that the entropy behaves similarly as in $\mathcal{N}=4$ SYM
discussed above, \cf\ the inset in
fig.~\ref{fig:F_and_U_N4_varT}. While $\SQQ$ vanishes for $L=0$, it
increases monotonically for increasing $L$ towards its maximum at the
screening distance.
The internal energy $\UQQ(L)$ in the non-conformal models has a shape
similar to that in $\mathcal{N}=4$ SYM. In particular, its slope
increases towards the screening distance $\Ls$, too, which can be
traced back to the strong increase of $\SQQ$ towards $\Ls$.

For small distances $L$, the behavior of the internal energy is
dominated by that of the free energy because the entropy approaches
zero in all our models. Accordingly, like the free energy discussed in
sec.~\ref{sec:invest-non-conf}, also the internal energy in all
non-conformal models converges to one universal curve for small $L$,
namely the one in $\mathcal{N}=4$ SYM given by the zero-temperature
potential $\VQQ$ in \eqref{eq:77}.

Differences between the behavior of the internal energy in our
non-conformal models and the behavior in $\mathcal{N}=4$ SYM generally
appear at intermediate and large $L$. In all of our non-conformal models 
the entropy at fixed distance decreases relative to its value
in $\mathcal{N}=4$ SYM for the chosen degree of non-conformality 
(and the same holds for smaller values of the non-conformality parameter, see below).  
In the overall effect on the internal energy, however, the increase in
the free energy that we have seen in sec.~\ref{sec:invest-non-conf}
overwhelms the decrease in the entropy in $\UQQ=\FQQ+T\SQQ$,
such that the internal energy in our non-conformal models is
larger than in $\mathcal{N}=4$ SYM. 
This indicates that $\UQQ$ in $\mathcal{N}=4$ SYM might possibly 
constitute a lower bound on the internal energy for a large class of
non-conformal theories. 
Looking at the different non-conformal models in more detail, 
we find that the 1-parameter Einstein-frame model is
very robust against non-conformal deformation, and both $\SQQ$ and
$\UQQ$ stay very close to their respective values in $\mathcal{N}=4$
SYM for all distances $L$. The effect of the non-conformality 
is larger and again similar in the \SWT{} and 1-parameter models. 

\begin{figure}[t]
  \centering
  \includegraphics[width=.8\textwidth]{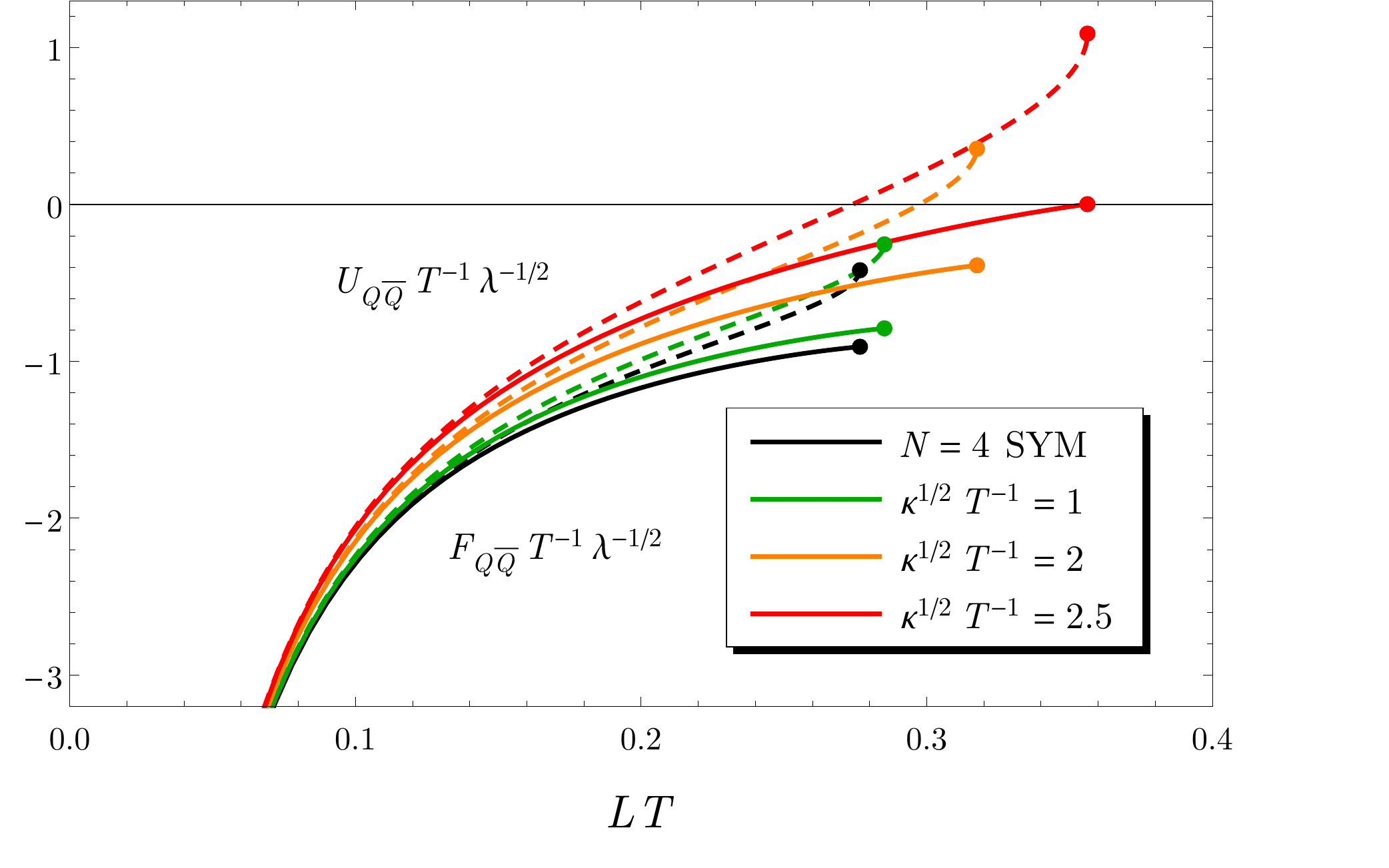}
  \caption{%
    Internal energy $\UQQ(L)/(T\sqrt\lambda)$ (dashed curves) and free
    energy $\FQQ(L)/(T\sqrt\lambda)$ (solid curves) at fixed
    temperature in $\mathcal{N}=4$ SYM (black curves) and the
    1-parameter string-frame model for varying deformation
    parameter. %
    For very small $L$, both $\UQQ$ and $\FQQ$ approach a universal 
    Coulombic curve.%
    \label{fig:F_and_U_N4_1pStr_varDef}
  }
\end{figure}

To not clutter the presentation, we have refrained in
fig.~\ref{fig:UQQ_all_models_SQQInset} from also showing the free
energy. To gain a better understanding of the relative behavior of the
free and the internal energy in a non-conformal model, we now focus on
the 1-parameter string-frame model as an example of a consistent
deformation of \AdSfive{}-Schwarzschild. Moreover, in this way we can
check whether the internal energy in the deformed model is larger than
the corresponding internal energy in $\mathcal{N}=4$ SYM also for
smaller values of the deformation.
In fig.~\ref{fig:F_and_U_N4_1pStr_varDef} we show the dependence of
the internal energy and free energy on the deformation parameter in
the 1-parameter string-frame model, starting from the undeformed
theory, \ie\ $\mathcal{N}=4$ SYM (black curves).
Like the free energy, the internal energy increases with increasing
deformation parameter, and indeed is larger than in $\mathcal{N}=4$
SYM for all choices of the deformation parameter.
As in $\mathcal{N}=4$ SYM, also in the non-conformal model the
internal energy approaches the free energy for small quark separation
$L$. 

We have performed a similar analysis for the other non-conformal models. 
We observe that the change of the internal energy with respect to $\mathcal{N}=4$ SYM
induced by non-conformality is positive. In the 1-parameter Einstein-frame 
model it is particularly small while it is larger and similar in the 
\SWT{} model and in the 1-parameter string-frame model. 
As the entropy approaches zero for small $L$ in all our
non-conformal models, \cf\ the inset in fig.~\ref{fig:UQQ_all_models_SQQInset}, 
we find that at small quark separations, $\FQQ$ and $\UQQ$ in
all theories approach as a common limit the free energy in
$\mathcal{N}=4$ SYM, which in turn approaches the zero-temperature
potential $\VQQ$ for small $L$. 
Thus, in all of our models the internal and free energy differ only
for intermediate and large quark separations. A similar behavior 
is also observed in lattice simulations \cite{Kaczmarek:2005gi}. 

\section{Free energy, entropy, and internal energy of a single heavy quark}
\label{sec:single-quark-free}

In the derivation of the holographic formula for the $\QbarQ$ free
energy in sec.~\ref{sec:general-discussion}, we have obtained as a
by-product the definition \eqref{eq:52} of the free energy of a
single quark in the hot medium described by our holographic models.
In this section, we will put that relation to use and compute the free
energy of a single heavy `test' quark. We will also study the entropy and
internal energy associated with the free energy.
An analysis of this set of single-quark quantities in our class of
non-conformal models, with a focus on the impact of non-conformal
deformations of $\mathcal{N}=4$ SYM, has not been performed in the
literature so far.%
\footnote{Previous work with different focus includes
  \cite{Finazzo:2014zga} which studies the single-quark free
  energy in a bottom-up framework tuned to model Yang--Mills
  thermodynamics, and the recent works 
  \cite{Iatrakis:2015sua,Fadafan:2015ynz}. 
  For a general discussion of
  single-quark thermodynamics in holographic models see
  \cite{Noronha:2010hb}.}

Let us first point out that the free energy of a single heavy 
quark is often defined in terms of the expectation value of a 
Polyakov loop. This definition is most frequently used in the 
imaginary-time formalism where inverse temperature is 
identified with the size of the compactified time direction. 
In our intrinsically real-time formalism, the definition of the 
single-quark free energy in terms of a Wilson line as given 
in section \ref{sec:general-discussion} is more natural. 
For a general discussion of the relation of the single-quark 
free energy to Polyakov loops in holography see 
\cite{Noronha:2009ud,Noronha:2010hb}. Our result 
for $F_Q$ in $\mathcal{N}=4$ SYM (see below), for example, 
coincides with the result obtained in \cite{Noronha:2010hb}. 

We have defined the free energy $F_Q$ in \eqref{eq:52}, and
define the corresponding entropy and internal energy by 
standard thermodynamic relations,
\begin{align}
  S_Q &= -\frac{\del F_Q}{\del T} \,,\label{eq:112}\\
  U_Q &= F_Q+TS_Q \,.\label{eq:113}
\end{align}

Let us start with $\mathcal{N}=4$ SYM. In this case,
the expression \eqref{eq:52} can easily be evaluated explicitly, using the
AdS-Schwarzschild metric \eqref{eq:78}. We obtain for the single-quark
free energy, entropy, and internal energy 
\begin{equation}
  \label{eq:72}
  F_Q = -\frac{\sqrt\lambda}{2}\,T\,,\qquad S_Q = \frac{\sqrt\lambda}{2}\,,
  \qquad U_Q = 0 \,.
\end{equation}
These values have also been obtained in \cite{Noronha:2010hb}. 
Since $\mathcal{N}=4$ SYM is a conformal theory, at $T>0$ only the
temperature itself is available as a dimensionful quantity for the
problem at hand. Thus, the relation $F_Q\propto T$ already follows
from dimensional analysis. The non-analytic square-root dependence on
the 't Hooft coupling is, however, a non-trivial outcome of the
computation. Interestingly, the free energy is entirely determined by
the entropic contribution, and the internal energy vanishes.

We can analytically compute the above quantities also in the \SWT\
model and the 1-parameter Einstein-frame model.
Let us start with the \SWT\ model. Due to the simple relation of the
temperature and the horizon position, $\zh=1/(\pi T)$, we can
explicitly express $F_Q$ as a function of $T$ and compute the entropy
and internal energy. We find
\begin{align}
  F_Q &= -\frac{\sqrt\lambda}{2\pi}\exp\left(\frac{c^2}{\pi^2T^2}\right)
  \left[\pi T - 2c\operatorname{\mathcal{F}}\left(\frac{c}{\pi T}\right)\right]
  \,,\label{eq:73}\\
  S_Q &= \frac{\sqrt\lambda}{2}\exp\left(\frac{c^2}{\pi^2T^2}\right) \,,\label{eq:74}\\
  U_Q &= \frac{\sqrt\lambda}{2\sqrt\pi}\,c\erfi\left(\frac{c}{\pi T}\right)\,,\label{eq:191}
\end{align}
where $\operatorname{\mathcal{F}}$ is the Dawson integral and $\erfi$
the `imaginary' error function defined by $\erfi(x)=-\i\erf(\i x)$.%
\footnote{The Dawson integral is defined by
  \begin{equation*}
    \operatorname{\mathcal{F}}(x)=\e^{-x^2}\int_0^x\d{}y\,\e^{y^2} \,,
  \end{equation*}
  and the error function by
  \begin{equation*}
    \erf(x) = \frac{2}{\sqrt{\pi}} \int_0^x \d t\, \e^{-t^2} \,,
  \end{equation*}
  see \cite{Abramowitz}. The latter is related to the Dawson
  integral by
  $\operatorname{\mathcal{F}}(x)=-\i \sqrt\pi\e^{-x^2}\erf(\i x)/2$.}
We recall that $\lambda$ denotes the bulk quantity defined by
$\sqrt\lambda=\LAdS^2/\alpha'$ and is a proxy for the coupling
strength in the boundary theory, \cf~\eqref{eq:190} and the
discussion thereof. An expression for $F_Q$ equivalent to 
\eqref{eq:73} had been found in \cite{Andreev:2009zk}. 
As we will see explicitly below, for high
temperatures $F_Q$, $S_Q$, and $U_Q$ approach their values in
$\mathcal{N}=4$ SYM from above. However, as $T$ is lowered they
significantly increase above the conformal values.

In the 1-parameter Einstein-frame model, we can only find closed-form
expressions for $F_Q$, $S_Q$, and $U_Q$ as functions of $\zh$, since we
cannot analytically invert the temperature function $T(\zh)$
in \eqref{eq:1}. We find
\begin{align}
  F_Q(\zh) &= -\frac{\sqrt\lambda}{8\sqrt{2}\pi}\sqrt\kappa
  \left[4\Gammafct{\frac{3}{4}}
    + \operatorname{\gamma}\left(-\frac{1}{4},\frac{\kappa^2\zh^4}{4}\right)\right]
  \,,\label{eq:75}\\
  S_Q(\zh) &= \frac{2\sqrt\lambda}{3}\,\frac{1}{\kappa^2\zh^4}\,
  \frac{\exp\left(-\frac{1}{2}\kappa^2\zh^4\right)\left[\exp\left(\frac{1}{4}\kappa^2\zh^4\right)-1\right]^2}{1-\exp\left(\frac{1}{4}\kappa^2\zh^4\right)+\frac{1}{3}\kappa^2\zh^4}\,,\label{eq:76}\\
  U_Q(\zh) &= \frac{\sqrt\lambda}{6\pi}\,
             \left\{\frac{1}{\zh}\,\frac{1-\exp\left(-\frac{1}{4}\kappa^2\zh^4\right)}
             {1-\exp\left(\frac{1}{4}\kappa^2\zh^4\right)+\frac{1}{3}\kappa^2\zh^4}
  - \frac{3\sqrt{2}}{8}\sqrt\kappa\left[4\Gammafct{\frac{3}{4}}
             +\operatorname{\gamma}\left(-\frac{1}{4},\frac{\kappa^2\zh^4}{4}\right)\right]\!\right\}\,,
\end{align}
where $\operatorname{\gamma}$ is the incomplete
$\operatorname{\Gamma}$-function.%
\footnote{The incomplete $\operatorname{\Gamma}$-function is defined by
  \begin{equation*}
    \operatorname{\gamma}(a,x) = \int_x^\infty\d{}t\,t^{a-1}\e^{-t} \,,
  \end{equation*}
  see \cite{Abramowitz}. It is related to the ordinary 
  $\operatorname{\Gamma}$-function by
  $\operatorname{\gamma}(a,0)=\Gammafct{a}$.}
Although we are not able to express these quantities symbolically in
terms of the temperature $T$, it is straightforward to analyze their
behavior numerically.

For the 1-parameter string-frame model, the additional terms in the
warp factors prohibit a solution of the integrals for $F_Q$, $S_Q$,
and $U_Q$ in closed form. Nevertheless, we can numerically evaluate
the definition \eqref{eq:52} for the free energy and easily compute
the entropy and internal energy from it.

When we consider a single heavy quark there is no intrinsic length scale 
in the problem, and as a consequence we cannot express our results 
in terms of a dimensionless quantity similar to $L T$ that we used in the 
case of the quark--antiquark pair before. Instead, the results will depend 
on the temperature in each of the different models. In order 
to sensibly compare temperatures in these different theories we choose, 
among various possibilities, the following procedure which was discussed for 
example in \cite{Panero:2009tv} in the context a specific holographic model 
\cite{Gursoy:2010fj}. We consider the 
dimensionless trace of the energy--momentum tensor, $(\epsilon-3p)/T^4$, 
which was calculated in \cite{Schade:2012ah} for our non-conformal models. 
It turns out to have a pronounced maximum, the position of which is a 
function of the respective deformation parameter. At temperatures above 
the position of the maximum, the behavior of $(\epsilon-3p)/T^4$  
strongly resembles that found above $\Tc$ in lattice studies of QCD, 
for example in \cite{Cheng:2007jq,Bazavov:2009zn,Panero:2009tv}. 
We therefore identify the position of the maximum found in our models 
with $\Tc$ \cite{Schade:2012ah}, which gives $\Tc/c\approx 0.494$ and 
$\Tc/\sqrt\kappa\approx 0.394$ in the \SWT\ model and in the 
1-parameter models, respectively. For definiteness, we assume $\Tc=176\MeV$ 
to introduce physical units, but the choice of this particular value will 
not be relevant for our results.\footnote{Various other methods 
to compare the temperatures of different theories or to fix the parameters 
of a holographic theory for comparison to QCD have been discussed 
in the literature, see \cite{Andreev:2006eh,Kajantie:2006hv,Gubser:2006qh}
for some examples. 
The values for the deformation 
parameters that we obtain from fixing the position of the maximum in 
$(\epsilon-3p)/T^4$ to a particular value of $\Tc$ are in the same ballpark 
as those obtained with other methods, see for instance 
\cite{Andreev:2006eh,Kajantie:2006hv}. 
We would like to point out that the qualitative conclusions below 
do not change if we vary the deformation parameters 
around the values given here.} 
We emphasize that our non-conformal models do not exhibit an actual 
phase transition and hence do not have a `critical temperature'. The models 
are expected to resemble the high-temperature (deconfined) phase of QCD. 
With the procedure just described we only fix the temperature range in 
which our models should be compared to QCD, $T \ge \Tc$. Accordingly, 
we will show our observables only in this range. 
Finally, in $\mathcal{N}=4$ SYM the trace of the
energy--momentum tensor vanishes identically for all temperatures, so that 
there is no analogous way to define $\Tc$. Thus, the choice $\Tc=176\MeV$ is
completely arbitrary in this case. In the following, we scale 
$F_Q$, $S_Q$, and $U_Q$ such that the dependence on $T$ becomes 
trivial for $\mathcal{N}=4$ SYM. 

\begin{figure}[t]
  \centering
  \includegraphics[width=.8\textwidth]{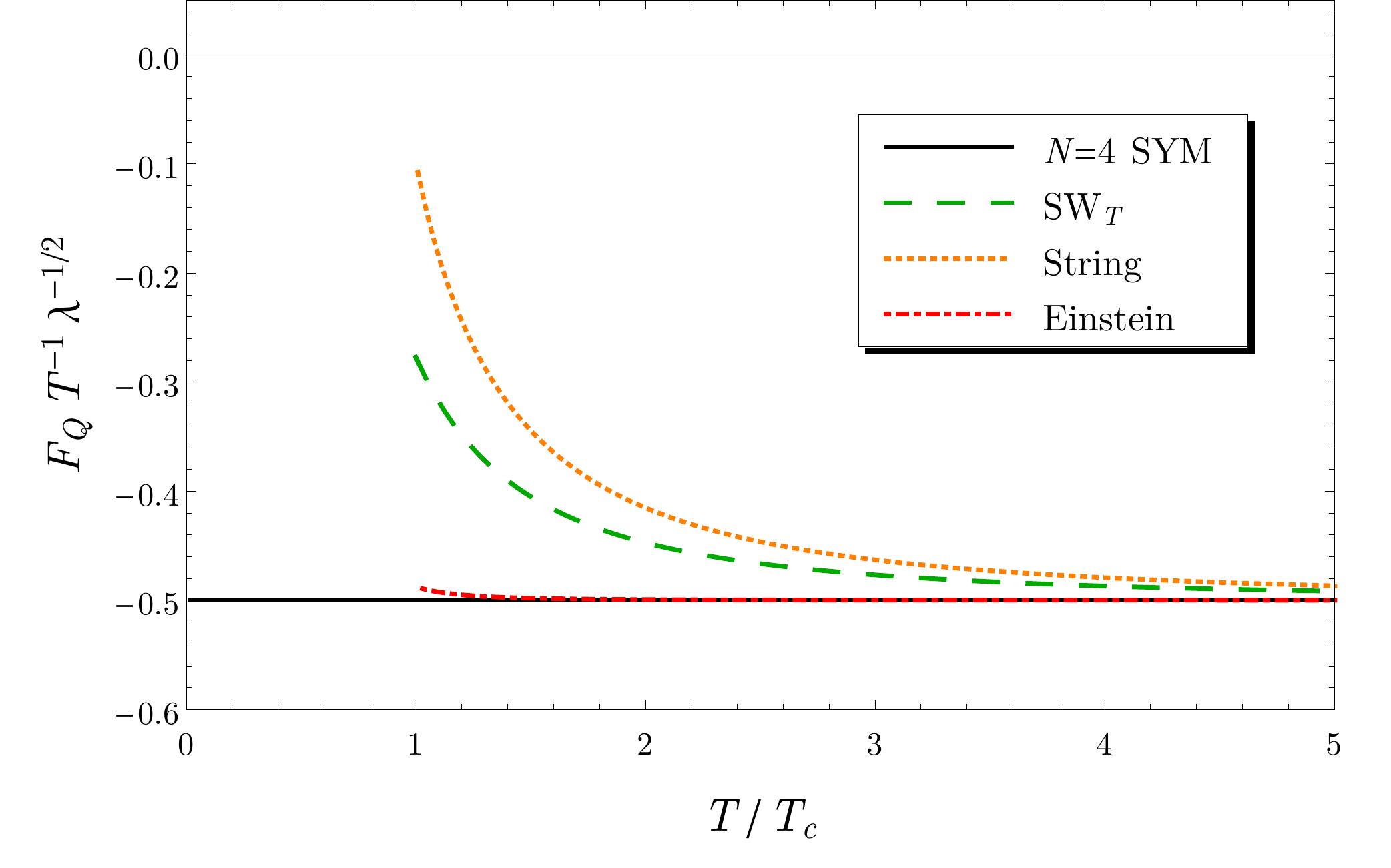}
  \caption{%
    Single-quark free energy as a function of temperature in
    $\mathcal{N}=4$ SYM and our non-conformal models. We have scaled
    out the dominant $T$-dependence of $F_Q$. The scale $\Tc$ is 
    explained in the text.%
    \label{fig:FQ_all_models_TOverTc}
  }
\end{figure}

\begin{figure}[t]
  \centering
  \includegraphics[width=.8\textwidth]{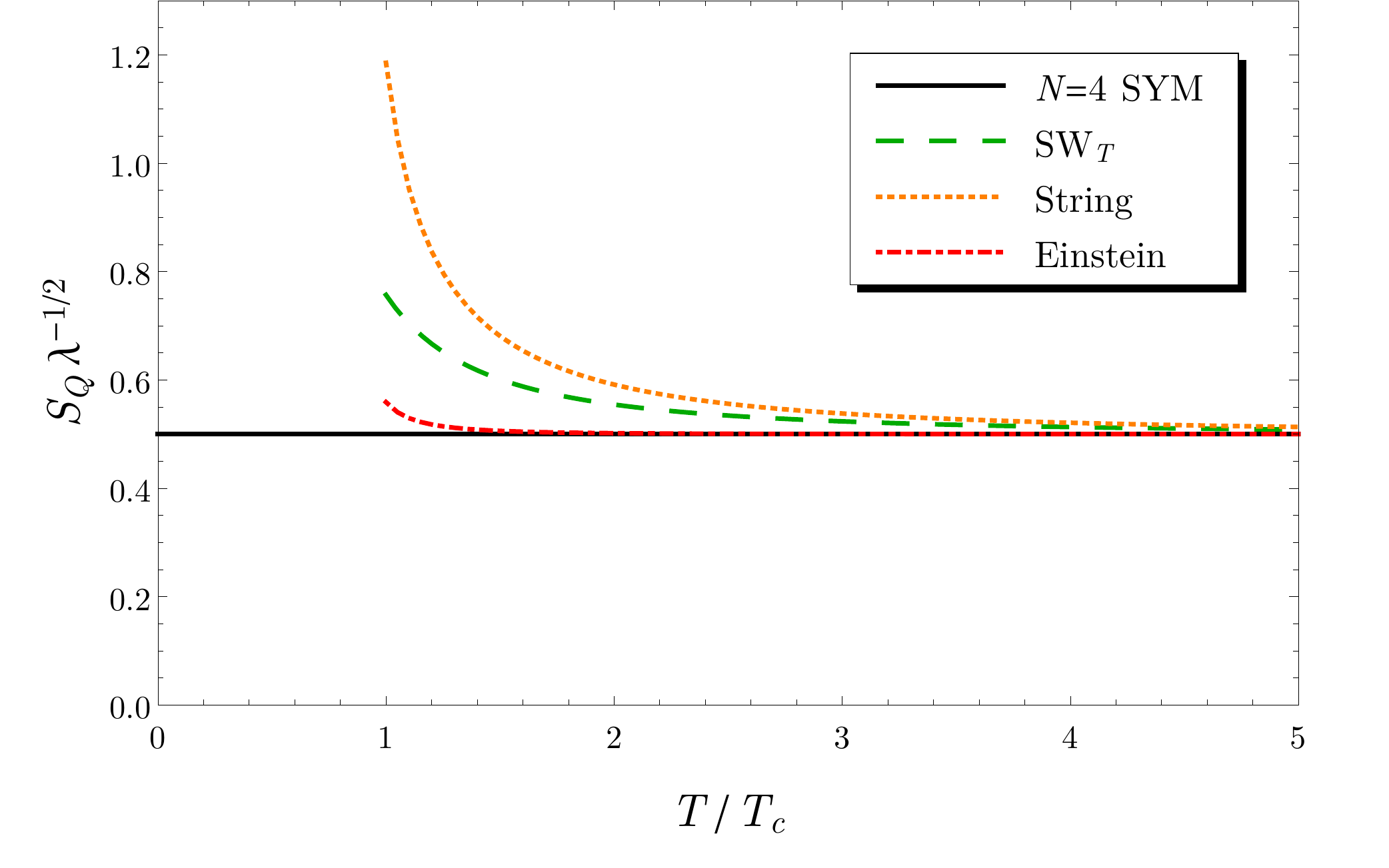}
  \caption{%
    Single-quark entropy as a function of temperature in
    $\mathcal{N}=4$ SYM and our non-conformal models. The scale $\Tc$ is 
    explained in the text.%
    \label{fig:SQ_all_models_TOverTc}
  }
\end{figure}

\begin{figure}[t]
  \centering
  \includegraphics[width=.8\textwidth]{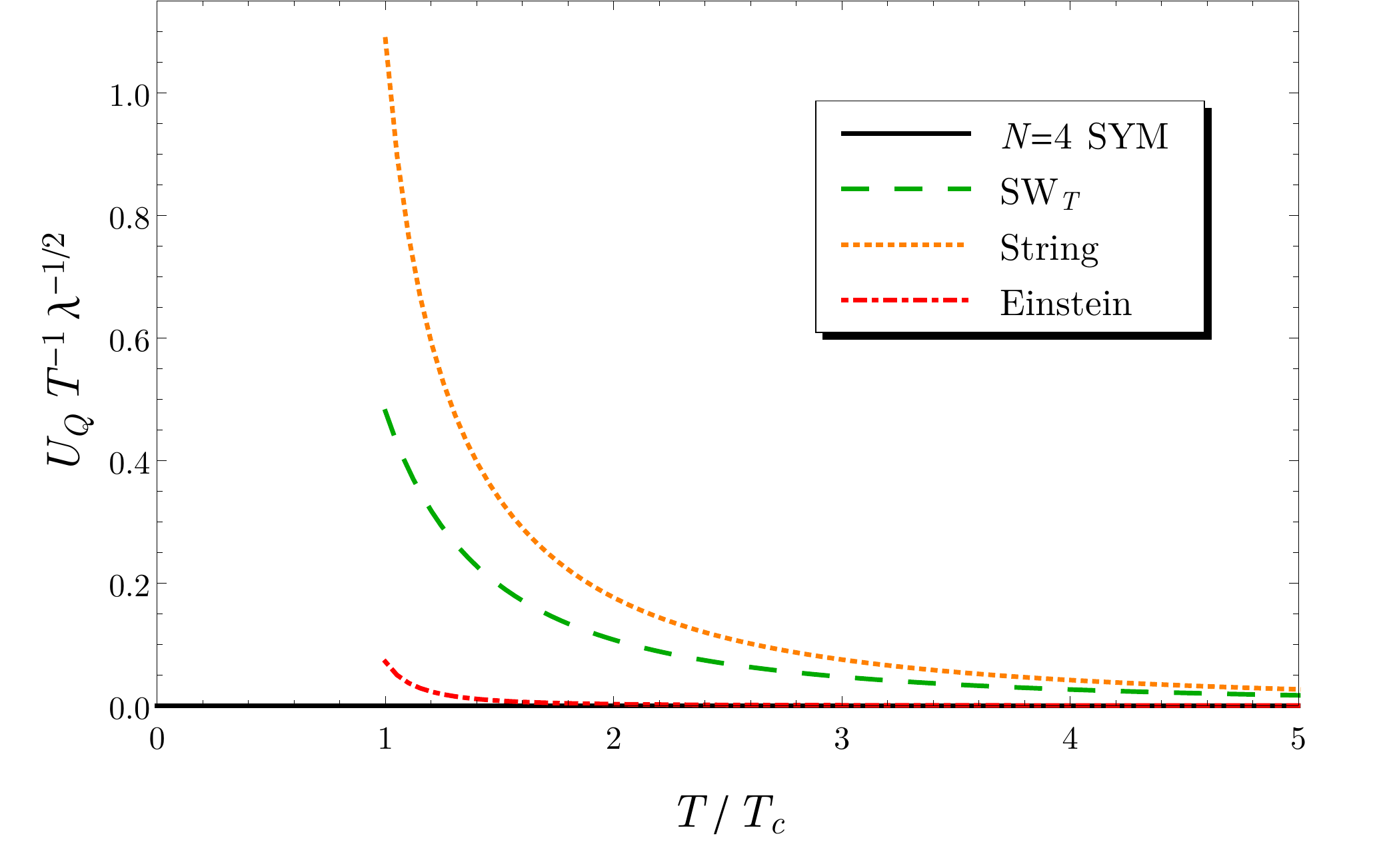}
  \caption{%
    Single-quark internal energy as a function of temperature in
    $\mathcal{N}=4$ SYM and our non-conformal models. 
    We have scaled out a factor of $T$. 
    The scale $\Tc$ is explained in the text. 
    Note that the internal energy vanishes identically in
    $\mathcal{N}=4$ SYM.%
    \label{fig:UQPhysUnits_all_models_TOverTc}
  }
\end{figure}

We plot the free energy, entropy and internal energy of the heavy quark 
as functions of the temperature in $\mathcal{N}=4$ SYM and in all our non-conformal
models in figs.~\ref{fig:FQ_all_models_TOverTc},
\ref{fig:SQ_all_models_TOverTc}, and
\ref{fig:UQPhysUnits_all_models_TOverTc}, respectively. 
We have seen in several observables before that the 1-parameter 
Einstein-frame model is very
robust and stays quantitatively close to $\mathcal{N}=4$ SYM. The
quantities $F_Q$, $S_Q$, and $U_Q$ are no exception and are very
close to their respective values in $\mathcal{N}=4$ SYM for almost all
$T$, as seen in the three figures.
All three quantities exhibit a stronger dependence on the temperature
in both the \SWT\ model and the 1-parameter string-frame model. The
latter deviates farthest from the behavior seen in $\mathcal{N}=4$
SYM.
For large temperatures, $F_Q$, $S_Q$, and $U_Q$ approach their values
in $\mathcal{N}=4$ SYM from above.  This implies an interesting
universal behavior. In all of our non-conformal models, $F_Q$, $S_Q$,
and $U_Q$ are larger than their respective values in $\mathcal{N}=4$
SYM for all temperatures. Even choosing a different procedure to
normalize the temperature scale in each model would not change this
finding. It appears likely that this observation holds in a large class 
of theories. 

\begin{figure}[t]
  \centering
  \includegraphics[width=.8\textwidth]{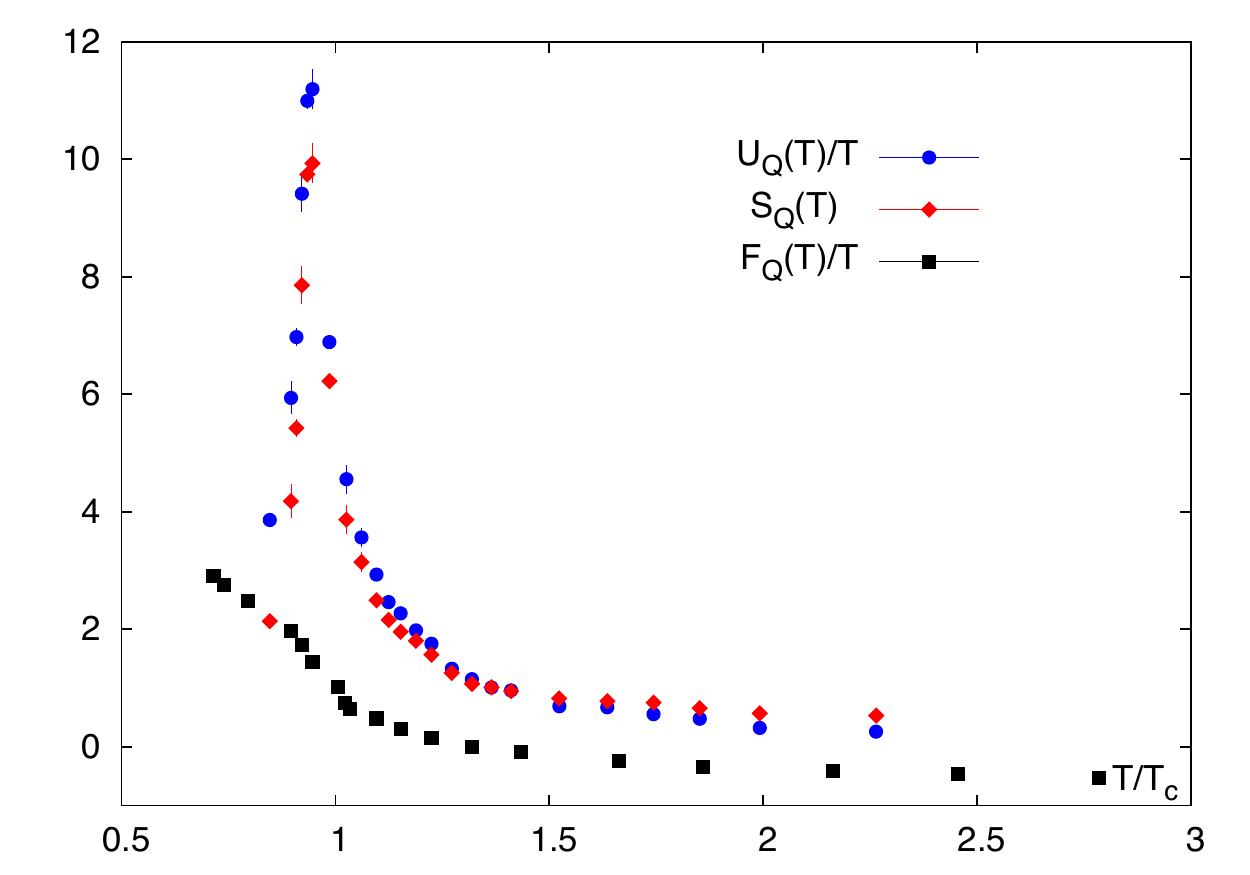}
  \caption{%
    Free energy over temperature ($F_Q/T$, black squares), 
    entropy ($S_Q$, red diamonds), and internal energy over temperature 
    ($U_Q/T$, blue circles) of a heavy quark 
    obtained from a 2+1-flavor lattice QCD calculation \cite{Kaczmarek:2007pb}. 
    \label{fig:F_and_S_inf_lattice}
  }
\end{figure}

Computations of the single-quark free energy in lattice QCD have been
performed for instance in
\cite{Kaczmarek:2002mc,Petreczky:2004pz,Kaczmarek:2005gi,Kaczmarek:2007pb,Bazavov:2013yv,Bazavov:2016uvm}.
In these studies, the single-quark free energy is defined in terms of the
large-distance behavior of the expectation value of a Polyakov loop
correlator.
The latter yields the free energy of a heavy $\QbarQ$ pair at large
quark separation, where the $\QbarQ$ free energy in fact approaches a
constant value $F_{\infty}$. At least in the deconfined phase where
the far-separated quarks are screened from each other, one can
interpret that free energy as twice the single-quark free energy,
$F_{\infty}=2F_Q$. In a similar way, lattice results have been 
obtained for the entropy $S_Q$ and the internal energy $U_Q$ 
of a single heavy quark. 
In the following, for concreteness, we will compare to 2+1-flavor 
lattice QCD results from \cite{Kaczmarek:2007pb}. The qualitative
behavior that we are going to compare to is similar in the other
lattice studies cited above.\footnote{The recent study \cite{Bazavov:2016uvm} 
with 2+1 dynamical flavors with physical masses finds deviations from 
previous calculations, most clearly visible in a less steeply falling 
entropy $S_Q$ around $\Tc$. These deviations may be attributed 
to quark mass dependences and the lack of a continuum extrapolation 
in previous results. On the level 
of the comparison with our generic holographic models, however, 
this difference is of minor relevance.} The data for $F_Q/T$, 
$S_Q$ and $U_Q/T$ from \cite{Kaczmarek:2007pb} 
are shown in fig.~\ref{fig:F_and_S_inf_lattice}. For this figure we made 
use of the freedom to add a constant to $F_Q$ and $U_Q$, and 
we have chosen this constant as $- 0.6\,\Tc$ relative to the data 
shown in \cite{Kaczmarek:2007pb}. 

In the case of the single heavy quark the comparison of our holographic 
to the lattice results is not completely straightforward. As we have seen, 
in the non-conformal models $F_Q$, $S_Q$ and $U_Q$ exhibit a universal 
rise above their values in $\mathcal{N}=4$ SYM, but their size 
strongly depends on the chosen non-conformal model, and for each 
model on the choice of the deformation parameter. Further, the 
free energy and the internal energy are only defined up to an arbitrary 
constant shift. This freedom could be fixed for the case of a 
$Q\bar{Q}$ pair by fixing the small-distance limit of the free energy 
$\FQQ$ to the vacuum potential $\VQQ$ of $\mathcal{N}=4$ SYM, 
see sec.~\ref{sec:invest-non-conf}. 
There is no analogue of such a condition 
in the case of a single quark. 
Finally, in all our models the results for all three 
quantities are proportional to $\sqrt{\lambda}$. As we have discussed 
at the end of sec.~\ref{sec:string-gener-metr}, especially for the 
non-conformal models $\lambda$ should be considered as a free 
parameter of the model, and it can even be different for different 
values of the non-conformal deformation parameter. In view of 
these caveats, we restrict the comparison with the lattice results 
to some qualitative features. Further detailed study in concrete models 
is needed to obtain a more complete picture of this comparison. 
Obviously, we again concentrate on the region $T \ge \Tc$ as 
our holographic models are expected to apply to the deconfined phase only. 

The free energy in units of temperature $F_Q/T$ from \cite{Kaczmarek:2007pb} 
is represented by the black squares in fig.~\ref{fig:F_and_S_inf_lattice}. 
Closely above $\Tc$ we observe a decrease of $F_Q/T$ which flattens above 
about $1.3 \,\Tc$ and approaches a constant.\footnote{In 
\cite{Bazavov:2016uvm} it is observed that this quantity rises again 
at very high temperatures. This can be attributed to the perturbative 
running of the strong coupling constant. We thus should not expect 
to see this effect in a holographic model that is asymptotically AdS 
in the UV.} 
The free energy in our non-conformal holographic models exhibits a similar
behavior above $\Tc$, as we can see in fig.~\ref{fig:FQ_all_models_TOverTc} 
which also shows the ratio $F_Q/T$. Note that $F_Q/T$ is a constant 
in $\mathcal{N}=4$ SYM. Hence the non-conformality is crucial for 
obtaining an $F_Q/T$ that resembles data from lattice QCD. 

The lattice data for $S_Q$ are shown as red diamonds in 
fig.~\ref{fig:F_and_S_inf_lattice}. $S_Q$ from the lattice is 
peaked at $T=\Tc$. 
The entropy is a $T$-independent constant in $\mathcal{N}=4$ SYM,
again illustrating the need to introduce non-conformality to model
QCD physics at $T\gtrsim \Tc$. Indeed, in our non-conformal models we
find an increase of $S_Q$ as $T$ is lowered towards $\Tc$,
qualitatively similar to the lattice results, see fig.~\ref{fig:SQ_all_models_TOverTc}. 
A heuristic explanation for the decrease of the entropy with
temperature above $\Tc$ might be as follows. As the temperature
increases, the screening length in the medium decreases and the quark
interacts with a smaller volume around it. Thus, its phase space and, 
accordingly, its entropy decreases (\cf\ a related discussion in
\cite{Petreczky:2004pz}).

The lattice data for $U_Q/T$ are shown as the filled blue circles 
in fig.~\ref{fig:F_and_S_inf_lattice}. 
As in the case of $S_Q$, they exhibit a strong peak at $T=\Tc$ 
and a subsequent decrease and flattening. A similar behavior 
of $U_Q/T$ is found in our non-conformal models, see 
fig.~\ref{fig:UQPhysUnits_all_models_TOverTc}. Again, the 
$\mathcal{N}=4$ SYM result does not show such a behavior. 

In summary, our comparison of the single-quark free energy, entropy 
and internal energy shows an agreement of at least some main properties of 
these quantities in the deconfined region $T \ge \Tc$ relevant
for the application of our holographic models. In particular, we 
observe that it is necessary to introduce non-conformality in the 
holographic model in order to achieve similarity to the lattice results.

\section{Summary and conclusions}
\label{sec:summary}

The free energy of a heavy quark--antiquark pair in a thermal medium
provides important information about the medium and its interaction
with color sources.
In this paper we have reconsidered the calculation of the free 
energy of the static pair in the framework of the AdS/CFT correspondence, 
where it is related to the action of a macroscopic string connecting 
the quark and the antiquark and hanging down into the holographic 
dimension. We have argued that the UV renormalization required 
for this action should not introduce a temperature dependence. 
With this condition, a consistent picture for the behavior 
of the free energy emerges. Applying a temperature-independent 
renormalization procedure we in fact find the free energy from AdS/CFT 
to be in qualitative agreement with data from lattice gauge theory. 
We observe that a temperature-dependent renormalization 
procedure widely used in the literature gives rise to the (negative) 
binding energy of the pair rather than to its free energy. As we 
have shown, the free energy and the binding energy have a markedly different 
dependence on temperature. 
We have then computed the entropy and the internal energy of 
the static quark--antiquark pair in the medium from the free energy 
and have discussed their properties. In order to obtain these 
observables it is essential to have the correct temperature dependence 
of the free energy. Finally, we have also computed the 
free energy, the entropy and the internal energy of a single 
heavy quark in the thermal medium, and have compared their 
behavior to that obtained in lattice gauge theory. 

We have performed these calculations in several holographic theories, 
starting with pure AdS${}_5$ space dual to $\mathcal{N}=4$ SYM. 
We have also considered deformed AdS-type spaces holographically 
dual to non-conformal deformations of $\mathcal{N}=4$ SYM. The 
latter theories are expected to share more properties with the 
actual quark--gluon plasma studied in heavy ion collisions than does 
conformal $\mathcal{N}=4$ SYM. Our aim in the present paper 
was explicitly not to investigate a particular holographic model for QCD. 
Instead, we have examined several classes of non-conformal 
models and have looked for universal behavior of our observables 
under non-conformal deformations. We have indeed found hints 
that, for any given distance, the free energy $\FQQ$ and the 
internal energy $\UQQ$ of a static quark 
pair in our non-conformal theories are consistently larger than their 
respective values in $\mathcal{N}=4$ SYM. The binding energy $\EQQ$, 
on the other hand, consistently decreases with respect to its value 
in $\mathcal{N}=4$ SYM under non-conformal deformation. 
We expect similar bounds to hold in larger classes of holographic theories. 
Also the free energy, entropy and the internal energy of single heavy 
quarks show a universal behavior under non-conformal deformations, 
being consistently larger than the respective $\mathcal{N}=4$ SYM 
values. For these single-quark observables, a non-conformality is 
even necessary in order to obtain a non-trivial behavior. 

Our considerations concerning the renormalization of the temporal 
Wegner--Wilson loop and the corresponding temperature dependence 
in holographic theories are very general. Here we have studied  
the effects of a temperature-independent renormalization procedure 
only for simple classes of holographic theories. It would obviously 
be interesting to extend these investigations to more complex plasmas, 
for example with chemical potential, and to more 
sophisticated holographic models for the actual quark--gluon plasma. 

The question how heavy quarkonia behave in the quark--gluon 
plasma is highly relevant for phenomenology, but difficult to answer 
for theory. We hope that the holographic calculation  
of the free energy, entropy and internal energy of a static quark pair 
in a strongly coupled medium can give some useful input 
in this context. In particular, the holographic results 
could be helpful for constructing the potential  
to be used in the Schr\"odinger equation describing heavy 
quarkonia in the medium. Also the discussion whether the dissociation 
of heavy quarkonia in the quark--gluon plasma is due to 
an entropic force (see for example \cite{Kharzeev:2014pha,Satz:2015jsa}) 
might benefit from our results concerning the temperature 
dependence of the entropy and internal energy of the pair 
at strong coupling. 

\begin{acknowledgments}
We would like to thank S.\ Gubser, J.\ Pawlowski, A.\ Rothkopf, S.\ Theisen, 
and P.\ Wittmer for helpful discussions. 
A.\,S.\ acknowledges support in the framework of the cooperation 
contract between the GSI Helmholtzzentrum f\"ur Schwerionenforschung 
and Heidelberg University.
\end{acknowledgments}

%
%
\appendix

\section{Computation of the entropy of a heavy quark pair}
\label{sec:deta-comp-qbarq}

Here we give details on the computation of the $\QbarQ$
entropy $\SQQ$ discussed in sec.~\ref{sec:heavy-quark-entropy}.

We use the basic thermodynamic formula $\SQQ=-\del\FQQ/\del T$ where it is
understood that the inter-quark distance $L$ is to be kept constant.
An implementation of this formula is not entirely straightforward. The
issue that arises is that the free energy $\FQQ$, as well as the
distance $L$, are only known as integrals in terms of the bulk length
scales $\zt$ and $\zh$, see \eqref{eq:51} and \eqref{eq:35},
respectively. Therefore, while these integrals can be readily computed
numerically (or even analytically for $\mathcal{N}=4$ SYM), the
differentiation with respect to the temperature $T$ while keeping the
distance $L$ constant is more involved.

In the following we use a notation in which a vertical bar with a subscripted variable
indicates that this variable is kept constant. We will suppress any
dependence on a possible deformation parameter. If present in the
model under consideration, the deformation parameter is always assumed
to be kept constant.

We first note that in $\mathcal{N}=4$ SYM and in the \SWT\ model 
the relation between $\zt$ and $T$ is one-to-one, namely $T=1/(\pi \zh)$. 
For the consistently deformed 1-parameter models this relation 
is modified but indeed remains one-to-one in the parameter range 
we consider in this paper, $0 \le \sqrt{\kappa}/T \le 2.5$. The relation 
ceases to be one-to-one only for $\sqrt{\kappa}/T > 2.94$, 
see also footnote \ref{foot4}. It is therefore straightforward to obtain 
$\zh=\zh(T)$ for all our models. 

We further recall from sec.~\ref{sec:string-gener-metr} that 
for a given inter-quark distance $L$ smaller than the screening 
distance $\Ls$ there are two different string solutions connecting 
the quark and the antiquark, with the one staying further away 
from the horizon $\zh$ being energetically favored. Hence for 
$0 \le L \le \Ls$ there are two branches of solutions, of which we 
consider mainly the energetically favored branch in this paper. In this branch 
$L$ monotonically increases with increasing $\zt$, while in 
the branch with the energetically disfavored configurations 
$L$ monotonically decreases with increasing $\zt$. 
Treating each of the branches separately, we can therefore 
invert $L(\zt,T)$ to obtain $\zt=\zt(L,T)$.
 
As a consequence of these considerations we can write  
\begin{equation}
  \label{eq:63}
  \left.\frac{\del\FQQ(\zt,\zh)}{\del T}\right\rvert_L
  = \left.\frac{\del\FQQ}{\del\zh}\right\rvert_{\zt}\frac{\del\zh}{\del T}
  + \left.\frac{\del\FQQ}{\del\zt}\right\rvert_{\zh}\left.\frac{\del\zt}{\del T}\right\rvert_L \,,
\end{equation}
where on $\del\zh/\del T$ we may omit the specification of the
variable that is to be kept constant because $\zh$ is actually a
function of $T$ only.
Next, we have to evaluate $\del\zt/\del T$ with $L$ kept
constant. Using
\begin{equation}
  \label{eq:64}
  0 \stackrel{!}{=} \d{}L = \left.\frac{\del L}{\del\zt}\right\rvert_{\zh}\d\zt
  + \left.\frac{\del L}{\del\zh}\right\rvert_{\zt}\d\zh
  = \left.\frac{\del L}{\del\zt}\right\rvert_{\zh}\d\zt
  + \left.\frac{\del L}{\del\zh}\right\rvert_{\zt}\frac{\del\zh}{\del T}\d{}T \,,
\end{equation}
we derive
\begin{equation}
  \label{eq:65}
  \left.\frac{\del\zt}{\del T}\right\rvert_L
  = -\left(\left.\frac{\del L}{\del\zt}\right\rvert_{\zh}\right)^{-1}
  \left.\frac{\del L}{\del\zh}\right\rvert_{\zt}
  \frac{\del\zh}{\del T} \,.
\end{equation}
Finally, we obtain
\begin{equation}
  \label{eq:66}
  \SQQ = -\!\left.\frac{\del\FQQ(\zt,\zh)}{\del T}\right\rvert_L
  = -\!\left(\left.\frac{\del\FQQ}{\del\zh}\right\rvert_{\zt}
    - \left.\frac{\del\FQQ}{\del\zt}\right\rvert_{\zh}
    \frac{\left.\frac{\del L}{\del\zh}\right\rvert_{\zt}}{\left.\frac{\del L}{\del\zt}\right\rvert_{\zh}}\right)\frac{\del\zh}{\del T} \,,
\end{equation}
which can be directly implemented on the basis of the numerical
routines (or analytic expressions) for $\FQQ=\FQQ(\zt,\zh)$ and
$L=L(\zt,\zh)$.


%
%

\providecommand{\href}[2]{#2}\begingroup\raggedright
\endgroup

\end{document}